\documentclass[preprintnumbers,superscriptaddress,nofootinbib,aps,prd,floatfix,longbibliography]{revtex4-1}
\usepackage{amsmath,amssymb,amsbsy,amstext, amsthm, simplewick}
\usepackage[dvipsnames]{xcolor}
\usepackage{graphicx}
\usepackage{amsfonts}
\usepackage{upgreek}
\usepackage[titletoc]{appendix}
\usepackage{setspace}
\usepackage{color}
\usepackage{comment}
\usepackage{xspace}
\usepackage[colorlinks=true,urlcolor=blue
,anchorcolor=blue,citecolor=blue,filecolor=blue,linkcolor=red,menucolor=blue
]{hyperref}

\newcommand{\newc}{\newcommand}
\newc{\fpi}{f_{\pi}}
\newc{\etap}{\eta^{\prime}}
\newc{\llll}{\langle\lambda\lambda\rangle}
\newc{\FFd}{F^a\tilde F^a}
\newc{\qbar}{{\overline q}}
\newc{\TR}{{\rm Tr}}
\newc{\Kahler}{K\"ahler }
\newc{\Zbb}{{\mathbb Z}}
\newc{\Rt}{{\mathbb R}^3}
\newc{\Rf}{{\mathbb R}^4}
\newc{\So}{{\mathbb S}^1}
\newc{\zt}{{\mathbb Z}_2}
\newc{\RtSo}{{\mathbb R}^3\times{\mathbb S}^1}
\newc{\scriminus}{{\cal I}^-}
\newc{\scriplus}{{\cal I}^+}
\newc{\mpl}{M_p}
\newc{\Ricci}{\mathcal{R}}
\newc{\bv}{\phi}
\newc{\calU}{{\cal U}}
\newc{\calK}{K}
\newc{\calUi}{{\cal U}^{-1}}
\newc{\calG}{{\cal G}}
\newc{\calO}{{\cal O}}
\newc{\calQ}{{\cal Q}}
\newc{\calOb}{{\cal O}^\dagger}
\newc{\hphi}{{\hat\phi}}

\newcommand{\Mpl}{M_\mathrm{Pl}}
\newcommand{\Mpc}{\mathrm{Mpc}}

\newcommand{\GeV}{\mathrm{GeV}}
\newcommand{\MeV}{\mathrm{MeV}}
\newcommand{\eV}{\mathrm{eV}}
\newcommand{\cm}{\mathrm{cm}}

\newcommand{\TRH}{T_\mathrm{RH}}

\newcommand{\beq}{\begin{equation}}
\newcommand{\eeq}{\end{equation}}
\newcommand{\ba}{\begin{align}}
\newcommand{\ea}{\end{align}}

\newcommand{\br}{\mathrm{BR}}

\DeclareMathOperator\erf{erf}
\DeclareMathOperator\Si{Si}
\DeclareMathOperator\Ci{Ci}
\newcommand{\Neff}{N_\mathrm{eff}}

\newcommand{\MEarth}{M_{\oplus}}

\newcommand{\aeq}{a_{\mathrm{eq}}}
\newcommand{\zeq}{z_{\mathrm{eq}}}
\newcommand{\Teq}{T_{\mathrm{eq}}}

\newcommand{\rhot}{\tilde{\rho}}
\newcommand{\p}{\phi}
\newcommand{\vp}{\varphi}
\newcommand{\Gammat}{\tilde{\Gamma}}
\newcommand{\dels}{\delta_\phi}
\newcommand{\delr}{\delta_r}
\newcommand{\delm}{\delta_a}

\newcommand{\ther}{\theta_r}

\newcommand{\thest}{\theta_\phi}
\newcommand{\thert}{\theta_r}
\newcommand{\themt}{\theta_a}
\newcommand{\gamt}{\Gamma_\phi}
\newcommand{\tilk}{k}
\newcommand{\rhost}{\rho_\phi}
\newcommand{\rhort}{\rho_r}
\newcommand{\rhomt}{\rho_a}
\newcommand{\kt}{k}
\newcommand{\arh}{a_{\mathrm{RH}}}
\newcommand{\krh}{k_{\mathrm{RH}}}

\newcommand{\ahor}{a_{\mathrm{hor}}}

\newcommand{\rvir}{r_{\mathrm{vir}}}
\newcommand{\Mvir}{M_{\mathrm{vir}}}
\newcommand{\vvir}{v_{\mathrm{vir}}}

\newcommand{\keq}{k_{\mathrm{eq}}}

\newcommand{\MRH}{M_{\mathrm{RH}}}
\newcommand{\class}{\texttt{CLASS}\xspace}
\newcommand{\cnad}{c_{\mathrm{nad}}}
\newcommand{\Hcon}{\mathcal{H}}
\newcommand{\kosc}{k_{\mathrm{osc}}}
\newcommand{\aosc}{a_{\mathrm{osc}}}
\newcommand{\Tosc}{T_{\mathrm{osc}}}
\newcommand{\zcoll}{z_{\mathrm{c}}}
\newcommand{\lcdm}{\Lambda\mathrm{CDM}}

\newcommand{\tauenc}{\tau_{\mathrm{enc}}}
\newcommand{\Mosc}{M_{\mathrm{osc}}}

\newcommand{\vrel}{v_\mathrm{rel}}

\newcommand{\pc}{\mathrm{pc}}
\newcommand{\kcompt}{k_\mathrm{Compt}}
\newc{\ga}{g_{a\gamma\gamma}}
\newc{\msun}{M_{\odot}}

\begin{document}
\title{Imprints of the Early Universe on Axion Dark Matter Substructure}

\begin{abstract}
  Despite considerable experimental progress large parts of the axion-like particle (ALP) parameter space remain difficult to probe in terrestrial experiments. 
  In some cases, however, small-scale structure of the ALP dark matter (DM) distribution is strongly enhanced, offering opportunities for astrophysical tests. 
  Such an enhancement can be produced by a period of pre-nucleosynthesis early matter domination (EMD).
This cosmology arises in many ultraviolet completions and generates the correct relic abundance for weak coupling $f_a\sim 10^{16}$ GeV,  ALP masses in the range $10^{-13}$ eV $<m_a < 1$ eV, and without fine-tuning of the initial misalignment angle. This range includes the QCD axion around $10^{-9}-10^{-8}$ eV. 
EMD enhances the growth of ALP small-scale structure, leading to the formation of dense ALP miniclusters which can contain nearly all of 
DM (depending on ALP mass and reheating temperature). 
We study the interplay between the initial ALP oscillation, reheating temperature, and effective pressure to provide analytic estimates of 
the minicluster abundance and properties. ALP miniclusters in the EMD cosmology are denser and more abundant than in $\Lambda\textrm{CDM}$.
While enhanced substructure generically reduces the prospects of direct detection experiments, 
we show that pulsar timing and lensing observations can discover these minihalos  over a large range of ALP  masses and reheating temperatures.
\end{abstract}

\author{Nikita Blinov}
\affiliation{Fermi National Accelerator Laboratory, Batavia, IL 60510, USA\\[0.1cm]}
\affiliation{Kavli Institute for Cosmological Physics, University of Chicago, Chicago, IL 60637, USA\\[0.1cm]}

\author{Matthew J. Dolan}
\affiliation{ARC Centre of Excellence for Particle Physics at the
  Terascale, School of Physics, University of Melbourne, 3010, Australia\\[0.1cm]}

\author{Patrick Draper}
\affiliation{Department of Physics, University of Illinois, Urbana, IL 61801, USA\\[0.1cm]}

\preprint{FERMILAB-PUB-19-560-A-T}
\date{\today}
\maketitle

\section{Introduction}
\label{sec:intro}

Axion-like particles (ALPs) provide a compelling and elegant explanation 
for the dark matter (DM) of the universe~\cite{Abbott:1982af,Dine:1982ah,Preskill:1982cy,Arias:2012az}. These DM candidates arise 
in ultraviolet (UV) completions of the Standard Model (SM) as pseudo-Nambu-Goldstone bosons 
of spontaneously broken global symmetries, or as zero modes of higher-dimensional gauge fields~\cite{Arias:2012az,Svrcek:2006yi,Arvanitaki:2009fg,Cicoli:2012sz}. 
As such, their masses can be naturally light. A relic density of ALPs can be produced via 
several mechanisms, including misalignment, thermal and inflationary production, and from the decays of heavier 
particles or topological defects. In many well-motivated cases 
several mechanisms can contribute. For reviews, see, for example, Refs.~\cite{Sikivie:2006ni,Marsh:2015xka}.

The nonthermal production of ALPs suggests that both their abundance and late-time 
distribution are sensitive to physics in the UV. This is in contrast to thermally-produced 
DM (such as weakly-interacting massive particles, or WIMPs), where the abundance only depends on 
processes at energies similar to the DM mass. This UV-sensitivity can open a window 
into the pre-nucleosynthesis universe, where few other probes are currently available.

\begin{figure*}[t!]
  \centering
  \includegraphics[width=0.47\textwidth]{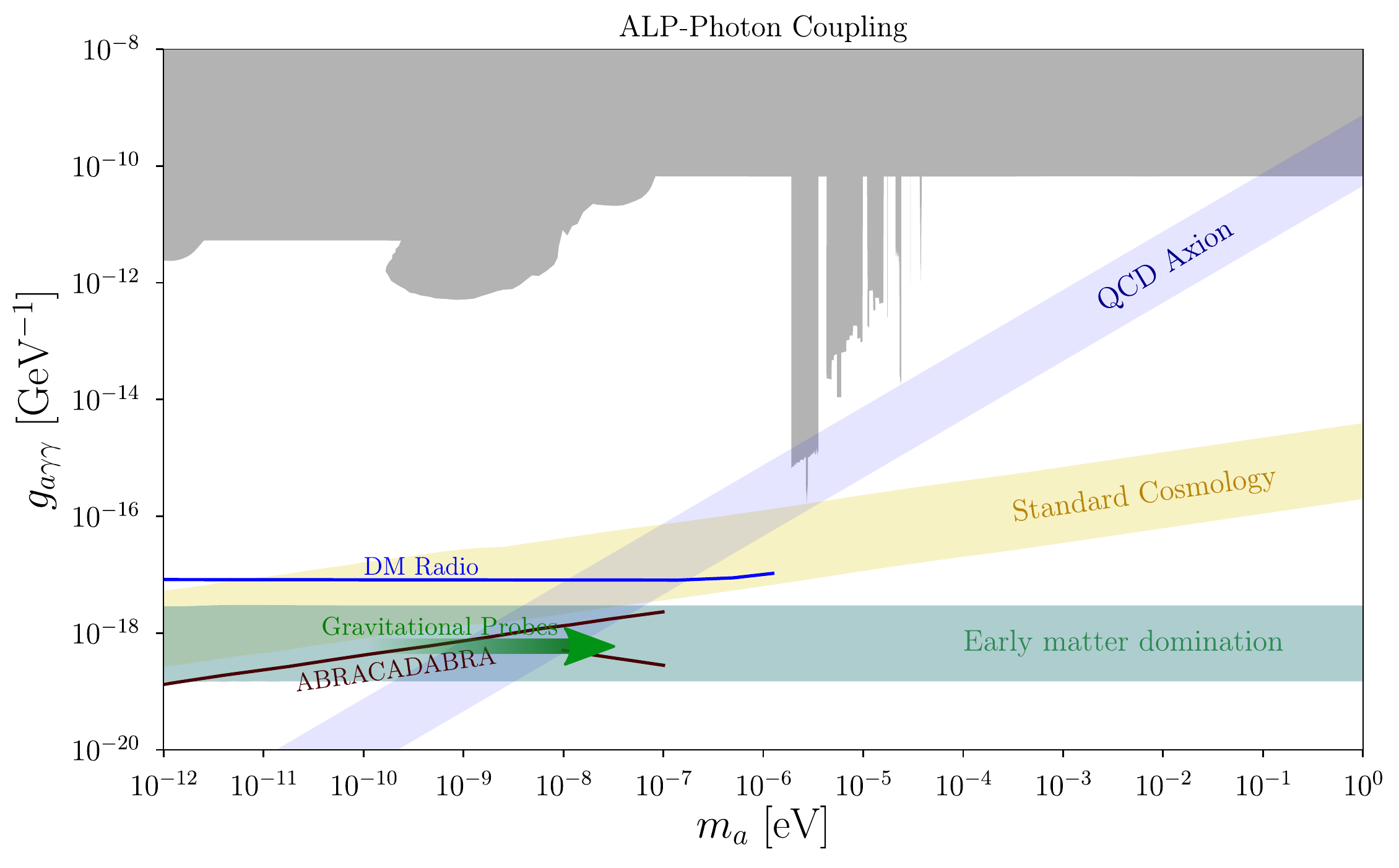}
  \includegraphics[width=0.47\textwidth]{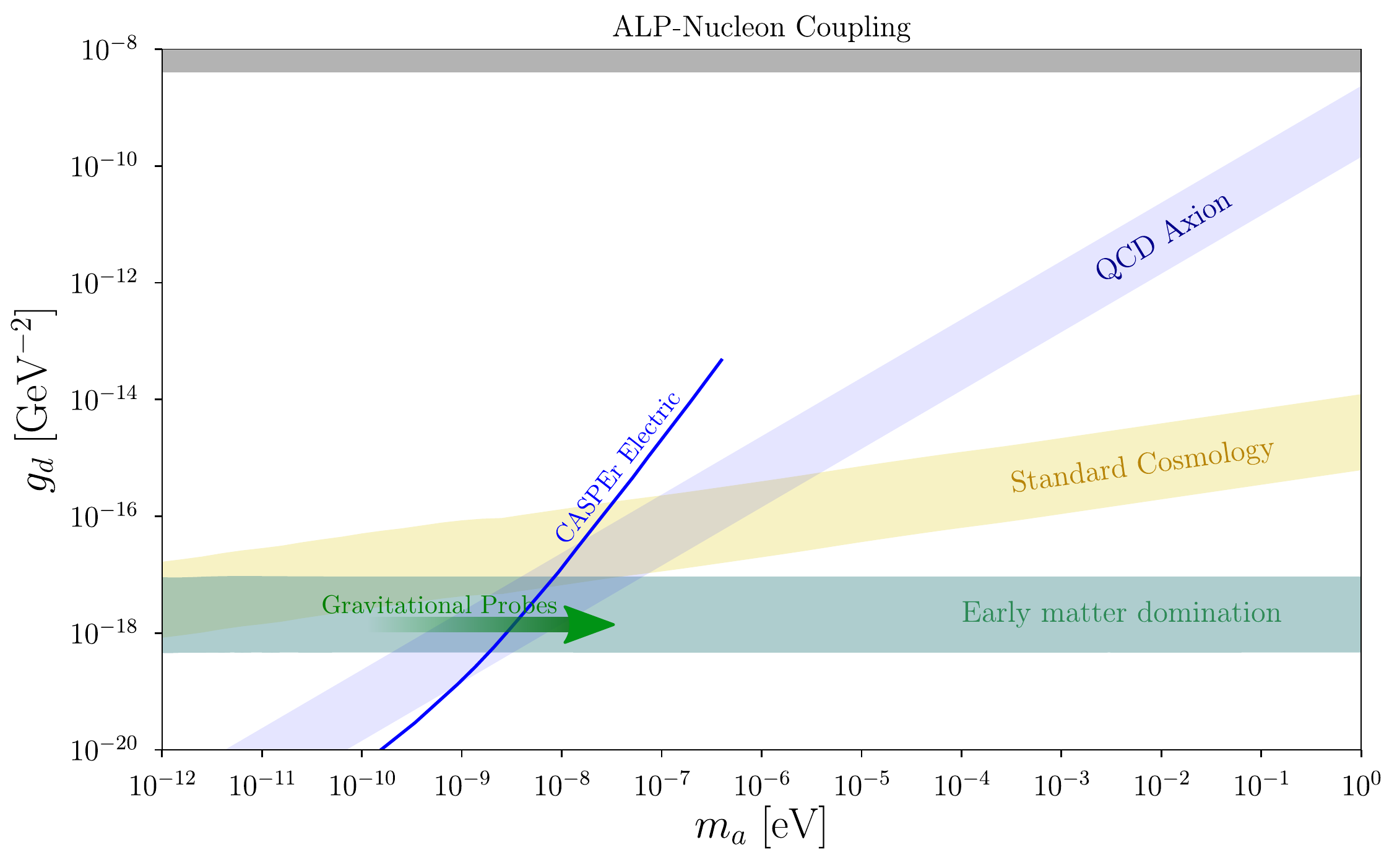}
  \caption{Existing constraints (gray regions), target parameter space (colored bands) and sensitivity of future experiments (colored lines) in the ALP mass $m_a$ and photon coupling $\ga \sim 1/f_a$ (left panel) and nucleon dipole moment coupling $g_d \sim 1/(f_a \Lambda_\mathrm{QCD})$ (right panel) planes. The QCD axion band is shown in blue and corresponds to a particular approximate mass-coupling relation, $m_a f_a \sim m_\pi f_\pi$. Relaxing this relationship but imposing saturation of the dark matter relic density results in other mass-coupling relations. These relations depend on the pre-BBN expansion history and initial misalignment angle $\theta_i$. The green and yellow bands correspond to cosmologies with early matter domination (with a reheat temperature of $10\;\MeV$) and standard radiation domination before nucleosynthesis and natural values of $\theta_i\sim 1$. Larger masses and lower couplings favored by EMD are challenging to probe in terrestrial experiments. In this paper we show that EMD enhances ALP DM small-scale structure, resulting in the formation of ALP minihalos.
These minihalos can be probed with lensing and pulsar timing observations through gravitational interactions alone for $m_a \gtrsim 10^{-10}\;\eV$ as indicated by the 
green arrow. We also show the far-future sensitivities of DM Radio~\cite{Chaudhuri:2014dla,Silva-Feaver:2016qhh} and ABRACADABRA~\cite{Kahn:2016aff,Ouellet:2018beu,Henning:2018ogd} to the ALP-photon coupling and CASPEr-Electric to the ALP-nucleon coupling~\cite{Graham:2013gfa,Budker:2013hfa,JacksonKimball:2017elr}. 
These experiments are projected to reach the EMD target region for a range of reheating temperatures (the upper and lower ABRACADABRA lines correspond to the broadband and resonant searches, respectively~\cite{Henning:2018ogd}). 
The left panel is adapted from Ref.~\cite{Blinov:2019rhb} including results of Ref.~\cite{Malyshev:2018rsh}.
\label{fig:masterplot}}
\end{figure*}

One of the principal means for discovering and measuring the properties of ALPs is through terrestrial direct detection experiments. Previously we have studied prospects for the direct detection of ALPs in a variety of cosmological scenarios~\cite{Blinov:2019rhb}, providing a range of experimental targets which are free of fine-tuning; some of these targets are shown in Fig.~\ref{fig:masterplot}.
The most difficult scenario to probe via direct detection involves a period of early matter domination (EMD). In EMD scenarios, the  value of the ALP decay constant required to achieve the correct DM relic density is large, corresponding to a small value of the ALP-photon coupling $\ga\sim 10^{-18}~\textrm{GeV}^{-1}$. This value is independent of the ALP mass over a large range of masses, including a range of QCD axion candidates around a nano-eV. 

Although challenging for ALP direct detection experiments, a period of EMD is an interesting possibility for early universe physics, motivated both by top-down model building and by bottom-up phenomenological considerations.

From the phenomenological perspective, the pre-nucleosynthesis expansion history has not been determined by observations, and it is natural to consider alternatives to the standard radiation-dominated assumption. Among these alternatives, a period of matter domination is perhaps the simplest possibility. Furthermore, as alluded to above, EMD is  a natural way to achieve the correct relic abundance of axions with decay constants around the GUT scale {\it independent of the axion mass}~\cite{banksdinegraesser,Blinov:2019rhb}. Whereas in the standard cosmology the axion fraction of the energy density grows linearly with temperature from $H
\sim m_a$ until matter-radiation equality, generally leading to overclosure for high-scale decay constants unless the axion is extremely light, in EMD this energy fraction is frozen around $(f_a/\Mpl)^2$ until reheating. The correct relic abundance is then obtained for $\Teq/\TRH\sim (f_a/\Mpl)^2$, independent of $m_a$.

From the top-down perspective,  EMD is thought to occur generically in string models of light axions~\cite{banksdinegraesser}, 
where  axions typically have GUT-scale decay constants~\cite{Svrcek:2006yi} and are accompanied by a saxion (a heavier scalar modulus partner of the axion) that comes to dominate the energy density at early times. 
Saxion moduli in these models typically have Planck-suppressed couplings, and there can also be other modulus fields which are similarly weakly-coupled (see, for example, the textbook discussion of Refs.~\cite{Polchinski:1998rq,Polchinski:1998rr} or the models discussed in Refs.~\cite{Svrcek:2006yi,Douglas:2006es,Acharya:2012tw}). 
Moduli fields are thus naturally long-lived, and the energy  stored in their oscillations can come to dominate the energy density of the early universe before they decay, leading to a period of EMD. If the moduli decay happens too late, the energy injection can ruin the success of Big Bang Nucleosynthesis (BBN): this is known as the cosmological moduli problem~\cite{Coughlan:1983ci,Banks:1993en,deCarlos:1993wie}. The moduli problem can be avoided if the moduli mass scale is above $\mathcal{O}(10)$~TeV, leading to reheating (RH) above 5 MeV and satisfying BBN constraints~\cite{Kawasaki:2000en,Hannestad:2004px,deSalas:2015glj,Hasegawa:2019jsa}. 

A period of EMD can modify dark matter physics in various ways. In WIMP scenarios, dark matter can be produced in the decay of the fields responsible for EMD, favoring different parts of the supersymmetric parameter space~\cite{Moroi:1999zb,Acharya:2008bk,Bose:2013fqa,Fan:2013faa,Blinov:2014nla}. 
In ALP scenarios, axions that begin to oscillate during an EMD phase have modified relic densities and perturbation growth relative to the standard radiation dominated cosmology. 
Ref.~\cite{Banks:1996ea} performed an early study of axion dark matter with a period of EMD; more recent studies include Refs.~\cite{Visinelli:2009kt,Nelson:2018via,Visinelli:2018wza,Ramberg:2019dgi} and projections for a wide range of proposed experiments for this scenario were given in Ref.~\cite{Blinov:2019rhb}.

In this work we explore the impact of a stage of early matter domination on the growth of ALP density perturbations, which has also recently been considered in Refs.~\cite{Nelson:2018via,Visinelli:2018wza} for ALPs and in Refs.~\cite{Giudice:2000ex,Erickcek:2011us,Barenboim:2013gya,Fan:2014zua,Erickcek:2015bda} for WIMPs.
Density perturbations grow linearly with the scale factor during EMD, as opposed to logarithmically during radiation domination. 
Linear growth leads to enhanced structure on scales that 
enter the horizon before the end of EMD. These structures decouple from the Hubble flow and collapse 
at high redshifts, leading to the formation of ALP miniclusters or minihalos (we use these terms interchangeably)~\cite{Hogan:1988mp,Kolb:1993hw,Kolb:1993zz,Kolb:1994fi,Enander:2017ogx,Fairbairn:2017sil}. Galactic dark matter halos are then hierarchically assembled from 
these miniclusters. If the miniclusters survive the galactic assembly process, their presence can significantly 
alter the optimal DM search strategy. For example, the terrestrial minicluster encounter rate 
may be too low for effective direct detection searches~\cite{Tinyakov:2015cgg,Berezinsky:2014wya}. On the other hand, such compact structures 
can be searched for via lensing and pulsar timing. Future lensing~\cite{Kolb:1995bu,VanTilburg:2018ykj,Dai:2019lud} and pulsar-timing searches~\cite{Dror:2019twh} will be able to probe compact DM substructure at an unprecedented level.

The formation of ALP miniclusters is known to occur for ALP initial conditions generated by 
post-inflationary Pecci-Quinn (PQ) breaking, topological defect decay or when ALP self-interactions are important.
In this work, we focus on the complementary case where the initial 
ALP perturbations are small $\mathcal{O}\left(10^{-4}\right)$, in analogy to the pre-inflationary PQ breaking scenario for the QCD axion. 
We show that even in this case, miniclusters form due to the enhanced growth during EMD. Unlike the post-PQ 
breaking case, the growth of ALP perturbations can be treated analytically. Consequently the minicluster distributions can 
be characterized with mild assumptions about their survival probability.
We generalize and extend previous analyses of Refs.~\cite{Nelson:2018via,Visinelli:2018wza} by considering non-QCD ALPs, treating growth before and after matter-radiation equality, and studying the impact of the ALP mass, sound speed effects, reheating temperature, and initial conditions on the growth and distribution of miniclusters.

Recent numerical work~\cite{Buschmann:2019icd} suggests QCD axion miniclusters formed from post-inflationary PQ breaking in the standard cosmology may be too light to be relevant for lensing and pulsar timing searches. In contrast, in the ALP parameter space where EMD provides a  natural explanation for the relic density,  we find that these searches, particularly photometric microlensing, offer strong sensitivity. In EMD scenarios it is plausible that most of the present-day relic density is bound in miniclusters, greatly weakening direct detection prospects. These conclusions affect a large range of weakly-coupled ALP models, including a QCD axion with a mass around an neV.

This work is organized as follows. In Section~\ref{sec:relic_abundance} we discuss the ALP relic density and a model for the period of early matter domination. In Section~\ref{sec:boltzmann} we assemble the Boltzmann equations for the evolution of the background energy densities and the ALP perturbations. The physics involves a number of different scales: the ALP mass (and hence scale of oscillation), the comoving horizon size, the ALP Jeans scale, and the reheating scale. We carefully assess the impact of each scale on the growth of perturbations, and we present numerical results tracking the perturbations from early matter domination, through reheating and into standard radiation domination, and through standard matter-radiation equality. In Section~\ref{sec:halo_function} we use the Press-Schechter formalism to estimate the statistical distributions of the ALP miniclusters. We discuss our results in Section~\ref{sec:analysis}, and study their implications for the minicluster survival rate, direct detection, and pulsar timing and lensing searches.
We conclude in Section~\ref{sec:conclusion}.

\section{ALP Relic Abundance}
\label{sec:relic_abundance}
Our model consists of the Standard Model, an ALP field $a$ which will constitute the dark matter, and a heavy scalar field $\p$. The energy density in coherent $\p$ oscillations dominates the universe at early times and redshifts like matter, leading to a period of EMD.
The scalar field $a$ is minimally coupled to gravity with action
\beq
S \supset \int d^4 x \sqrt{-g}\left[\frac{1}{2}(\partial a)^2 - V(a) \right], \label{eq:alp_action}
\eeq
where for simplicity we consider a quadratic potential
\beq
V(a) = \frac{1}{2}m^2 a^2. \label{eq:potl}
\eeq

If the field $a$ is an axion-like particle then we expect its interactions with SM fields to be suppressed by factors of $1/f_a$, where $f_a$ is the ALP decay constant. While critical for direct detection, these interactions are not relevant for the early universe cosmology we study in this paper. We assume that the ALP relic density is set through the misalignment mechanism~\cite{Preskill:1982cy,Abbott:1982af,Dine:1982ah}, and that the ALP mass is independent of temperature.

Higher-order terms may also be present in the potential of Eq.~\ref{eq:potl}. These terms delay the onset of ALP oscillations for large initial values of $a$, and including them can have $\mathcal{O}(1)$ impact on the relic density. For $a/f_a\lesssim 1$ these effects are unimportant. Nonlinear terms in the ALP equation of motion can also lead to important effects on small scales, such as soliton-like configuration known as axitons or oscillons~\cite{Kolb:1993hw,Vaquero:2018tib,Buschmann:2019icd,Olle:2019kbo,Arvanitaki:2019rax}. 
The precise effect of self-interactions depends on the interaction terms and temperature dependence of the ALP mass. 
On the whole, the presence of axitons could provide additional substructure within the miniclusters we identify below and would be interesting to pursue further through dedicated numerical studies.

In a Friedmann-Robertson-Walker cosmology, the equation of motion for the ALP background is
\beq
\frac{d^2 a}{dt^2} + 3 H \frac{d a}{dt} + m^2_a a =0 \label{eq:alp_eom} \, ,
\eeq
where $H$ is the Hubble parameter.
The evolution of the ALP field depends on the distribution of initial conditions $\theta_i = a_i/f_a$. If the ALP exists prior to inflation then $\theta_i$ is uniform across the initially causally-disconnected regions constituting the universe at the present time. This is the scenario we study in this paper.
Another possibility is that $\theta_i$ is stochastically distributed over the separate causal patches throughout the universe. In models where the ALP is a pseudo-Nambu Goldstone boson, this corresponds to the scenario where associated global symmetry is broken after inflation. This can be studied analytically using cosmological perturbation theory by taking an effective average misalignment angle corresponding to $\theta_i = \pi/\sqrt{3}$~\cite{Kolb:1990vq}. However, one also expects the formation of topological defects such as strings and domain walls at the boundaries of different causal patches which are not captured by this approach. The decay of these defects leads to large fluctuations in the ALP field which later evolve into miniclusters. Structure formation in this scenario has recently been studied numerically in Refs.~\cite{Vaquero:2018tib,Buschmann:2019icd} and analytically in Refs.~\cite{Fairbairn:2017dmf,Fairbairn:2017sil,Enander:2017ogx} using the Press-Schechter formalism, and leads to the formation of miniclusters~\cite{Hogan:1988mp,Kolb:1993hw,Kolb:1993zz,Kolb:1994fi}.

It is well known that the pre-inflationary PQ-breaking axion scenario generates isocurvature perturbations, which are strongly constrained by cosmic microwave background (CMB) measurements~\cite{Axenides:1983hj,Seckel:1985tj,Efstathiou:1986pba}. Suppressing these modes either implies an upper bound on the scale of inflation $H_I$~\cite{Fox:2004kb} or requires nontrivial axion-inflaton dynamics (see, e.g., Ref.~\cite{Fox:2004kb}). In the former case we estimate in App.~\ref{sec:isocurvature} that $H_I\lesssim 10^{9-10}$~GeV depending on the reheating temperature. This is less constraining than for the QCD axion in a standard cosmology~\cite{Visinelli:2009kt}. For rest of this paper we assume that inflation has taken place at sufficiently low scale to satisfy the isocurvature constraint.

The scalar field $\p$ comes to dominate the energy density in the early universe before it decays. Reheating occurs when the Hubble parameter is approximately equal to the decay width of the scalar, $H\sim \Gamma_\p$. During matter-domination $H\propto a^{-3/2}$, and the scale factor $a$ at reheating is approximately $\arh \sim \Gamma_\p^{-2/3}$.\footnote{We use $a$ to refer to both and ALP field and the cosmological scale factor. It will be clear from context which one is in use.} We denote the temperature of the universe when reheating occurs by $\TRH$. This is constrained by Big Bang Nucleosynthesis to be larger than $\mathcal{O}(\rm{MeV})$. The lowest reheat temperature we consider in this work is 5~MeV~\cite{Kawasaki:2000en,Hannestad:2004px,deSalas:2015glj,Hasegawa:2019jsa}.

In UV-complete models the field $\phi$ could correspond to a saxion or modulus field. We assume that the $\p$ decays predominantly into Standard Model fields, corresponding to radiation in the early universe. It is also possible that $\p$ decays into ALPs. In that case the ALP relic density would be made up partly from a population due to misalignment and partly from a population due to $\p$ decay.
Whether this population behaves as matter or radiation depends on the relative mass 
of $\phi$ and $a$, and on $\TRH$. However, for $m_a < \eV$ and the low reheat temperatures we are interested in, ALPs produced from $\phi$ decays are still 
relativistic at matter-radiation equality (MRE). Therefore, they contribute to the total energy density of the universe as dark radiation and 
their abundance is constrained by the concordance of standard cosmology with observations of the CMB and light element abundances. 
These constraints are conveniently expressed as bounds on the effective number of relativistic degrees of freedom, $\Neff$.
In the instantaneous decay approximation we estimate that ALPs from $\phi$ decays would contribute
\beq
\Delta\Neff \lesssim \frac{4}{7}\left[\frac{\br(\phi\rightarrow a\,a)}{\br(\phi \rightarrow \mathrm{SM}\;\mathrm{SM})} \right]g_*(\TRH),
\eeq
where the upper bound arises from assuming $\TRH \lesssim 10\;\MeV$; higher $\TRH$ is more weakly constrained due to additional SM 
entropy injections which dilute the relativistic ALP contribution relative to the SM.
CMB and BBN limit $\Delta\Neff \lesssim 0.5$ at $95$ \% CL (see, e.g., Refs.~\cite{Pitrou:2018cgg,Aghanim:2018eyx,Blinov:2019gcj}), which translates into $\br(\phi\rightarrow a\,a) < 0.08$. 
In the absence of self-interactions, this ALP component does not contribute to the formation of ALP clumps, so we set $\br(\phi\rightarrow a\,a) =0$ throughout this work. We also note that topological defects such as domain walls and strings tend to be irrelevant in pre-inflationary ALP scenarios, since inflation smooths out inhomogeneities. Recent numerical studies suggest that the magnitude of the defect contribution is small even in the post-inflationary ALP scenario~\cite{Klaer:2017ond,Buschmann:2019icd,Vaquero:2018tib} (although significant uncertainty remains  -- see Refs.~\cite{Gorghetto:2018myk,Kawasaki:2018bzv,Buschmann:2019icd}).
We therefore assume that the relic density is determined entirely by production through misalignment.

In the early universe while $H > m_a$ the field $a$ is effectively frozen in its initial value. As the universe expands and Hubble decreases and becomes comparable  with the ALP mass $H \sim m_a$ the ALP field starts evolving at time $t_{\rm osc}$ and oscillates in its potential. 
After oscillations begin, the ALP energy density redshifts as matter,
\beq
\rho_a = \frac{1}{2}  m^2_a f_a^2 \theta_i^2 (a(t_{\rm osc})/a)^3 + \mathcal{O}(H^2/m_a^2) \, ,
\eeq
where the $a^{-3}$ is the redshifting of the energy density with the scale-factor.

The ALP density in the current epoch can be shown to be approximately~\cite{Blinov:2019rhb}
\beq \label{eq:Omega_EMD}
\Omega_a h^2 \, \simeq \, 0.12 \times \left(\frac{f_a \theta_i }{9\times 10^{14} \, {\rm GeV}}\right)^2  \times \left(\frac{\TRH}{10\, \MeV} \right) \, ,
\eeq
where $\theta_i = a_i/f_a$ is the initial ALP misalignment angle. This equation holds for temperature-independent ALP masses, and also assumes that the reheating temperature $\TRH$ is lower than the ALP oscillation temperature $\Tosc$. If this were not the case then ALP oscillations would commence during radiation domination, and EMD would not have any impact on dark matter physics and structure formation.  Notably, Eq.~(\ref{eq:Omega_EMD}) is independent of the  ALP mass: EMD is an efficient mechanism for preventing heavier ALPs from over-closing the universe without fine-tuning  the misalignment angle~\cite{Draper:2018tmh,Blinov:2019rhb}. Similarly, it achieves the correct relic abundance for weaker couplings than in the standard scenario. In the next section we study the evolution of inhomogeneities in the ALP field and 
the corresponding cosmological density and velocity perturbations.

\section{Growth of ALP Perturbations}
\label{sec:boltzmann}

In this section we trace the evolution of the ALP perturbations through EMD, reheating, and 
into standard radiation domination. The initial ALP perturbations are small 
for our choice of initial conditions, so their growth during these stages is well-described 
by linear perturbation theory. We determine the evolution of the background densities, then 
find the Boltzmann equations and initial conditions describing the perturbations in 
the EMD field, ALPs and radiation. We solve the Boltzmann system
numerically and discuss the growth of perturbations of different physical sizes. Finally, we use these results to construct approximate transfer functions, which will be employed in the following 
section to characterize minicluster formation and distribution in the non-linear regime.

Our analysis closely follows that of Ref.~\cite{Erickcek:2011us} 
for WIMP-like DM with EMD, so it is useful to highlight some differences from the 
WIMP case. First, since the ALP masses are low, if they are to account for the cold DM of the universe, 
they cannot be produced during reheating in the decays of the massive particles responsible for EMD. Second, ALPs only behave like matter 
after they begin to oscillate. This sets a characteristic minimum scale for enhanced structure growth, and 
it modifies the Boltzmann equations compared to the WIMP scenario. Finally, the ALP coupling to the SM is so weak 
that there are no annihilations or decays of the ALP during reheating~\cite{Fan:2014zua}.

\subsection{Background}
We model the background evolution of the universe as a three fluid system adapted from~\cite{Chung:1998rq,Erickcek:2011us}. 
This scenario is described by the following evolution equations for the energy densities of the scalar field $\rho_\p$, the ALP $\rho_a$, and SM radiation $\rho_r$:
\begin{subequations} 
\begin{align}
  \frac{d \rho_\phi}{dt} + 3 H \rho_\phi & = - \Gamma_\phi \rho_\phi \\
  \frac{d  \rho_r}{dt} + 4 H \rho_r &= + \Gamma_\phi \rho_\phi \\
  \frac{d \rho_a}{dt} + 3 H \rho_a &=0.
\end{align}
\label{eq:background_sys}%
\end{subequations}
We will follow the conventions of Ref.~\cite{Erickcek:2011us} and 
set the scale factor and Hubble parameter at an initial time $t_0$ to be $a(t_0) = 1$ and $H(t_0) = H_1$. We also define dimensionless variables
\begin{align}
  \rhot_i & = \rho_i/\rho_{\mathrm{crit},0} \\
  \Gammat_\phi & = \Gamma_\phi/H_1,
\label{eq:dimless_vars}
\end{align}
where $\rho_{\mathrm{crit},0} = 3\Mpl^2 H_1^2 /(8\pi)$. We work 
in these scaled units throughout and drop the tildes below for simplicity. 
Physical quantities are ratios of scales and the dependence on $H_1$ drops out.

Equations~\ref{eq:background_sys} are valid after the ALP starts to oscillate; 
before this era, the energy density must be obtained by solving the field equation~\ref{eq:alp_eom}.
We obtain the initial conditions for Eq.~\ref{eq:background_sys} by assuming that 
$\phi$ dominates the energy density at early times, that the dominant
component of radiation has been produced 
by $\phi$ decay alone (i.e. any primordial contribution 
has been diluted away), and that the universe is flat. These assumptions give 
\begin{align}
\rho_r(t_0) & \approx \frac{2}{5}\Gamma_\p \rho_\p(t_0)\\
\rho_a(t_0) & \approx \rho_a(t_0)\\
\rho_\phi(t_0) + \rho_a(t_0) +  \rho_r(t_0) & \approx 1,
\end{align}
where the last condition can be solved for $\rho_\p(t_0) \approx 1 - \mathcal{O}(\Gamma_\p)$. 
Since we are considering times well before the standard matter-radiation equality, we have $\rho_a(t_0) \ll \rho_{\phi,r}(t_0)$.

The background evolution, Eq.~\ref{eq:background_sys}, can be solved in the early-time limit:
\begin{subequations}
\begin{align}
  \rho_\phi(t) & \approx \rho_\phi(t_0) a^{-3}\\
  \rho_a(t)    & \approx \rho_a(t_0) a^{-3}\\
  \rho_r(t) & \approx \rho_r(t_0) a^{-3/2}.
\end{align}
\label{eq:background_sys_early_sol}%
\end{subequations}
The ALP and $\p$ redshift as matter, but the radiation density redshifts slower than the usual $a^{-4}$ due to the $\p$ decays. 
These solutions are approximately valid until reheating when $\Gamma_\phi / H(\arh) \sim 1$, which occurs when 
\beq
\arh \sim \Gamma_{\phi}^{-2/3}.
\label{eq:scale_factor_at_rh}
\eeq

\begin{figure}
  \centering
  \includegraphics[width=0.47\textwidth]{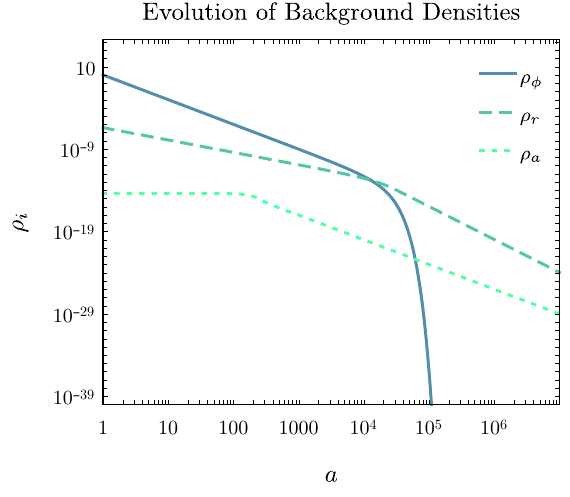}
  \caption{Evolution of background densities for a cosmology with early matter domination. At early times, the 
  energy budget of the universe is dominated by a non-relativistic field $\p$, which eventually decays into radiation (at 
  $a= \arh \approx 10^4$), 
  reheating the universe. The DM of the universe consists of an axion-like particle with density $\rho_a$. This density is constant before the ALP field commences oscillation at $a_{\rm{osc}} \approx 10^2$. We study 
  the growth of ALP perturbations in this background cosmology.
\label{fig:bg_density_evolution}}
\end{figure}

Compared to Ref.~\cite{Erickcek:2011us}, we assume in Eq.~\ref{eq:background_sys} 
that the DM component does not arise from decays of $\phi$ ($f=\br(\phi\rightarrow a\, a) = 0$ in their notation). 
Instead, the ALP initial condition is  set by requiring that we get the correct DM abundance by the time of 
(standard) matter-radiation equality. A representative numerical solution of the system in Eq.~\ref{eq:background_sys} is shown in Fig.~\ref{fig:bg_density_evolution}. 
Before it commences oscillating the axion energy density behaves as a component of dark energy and is constant. In this section we have presented the background evolution equations in terms of the time $t$. In all sections after this we will use either the scale factor $a$, or conformal time $\tau$ as our time variables.

\subsection{Perturbations and Initial Conditions}
\label{sec:pert_and_ic}


Density and velocity perturbations in the $\p$ and radiation fluids are governed by the Einstein and stress-energy conservation equations 
in the perturbed FRW space-time. The ALP fluid behaves as a cosmological 
constant before oscillations commence and as matter afterwards. 
Even deep in the matter-like regime, however, the ALP is 
not CDM-like at all scales. At small physical scales (large comoving wavevector $k$), the wave-like nature of the ALP 
introduces an effective pressure (and a corresponding sound speed) for the ALP fluid, preventing clustering below a characteristic 
scale, the Jeans wavenumber $k_J$, which is estimated below following Refs.~\cite{Hu:2000ke,Hwang:2009js,Arvanitaki:2009fg}. We 
focus on times after oscillations have begun and take into account these effects by 
integrating out the fast oscillations of the ALP, giving rise to effective fluid equations.
The derivation of these equations, including the 
ALP sound speed, is discussed in greater detail in Appendices~\ref{app:boltzmann} and~\ref{sec:ic_from_eom}. Here we collect the results. 
 
We work in the conformal Newtonian gauge and take the metric to be 
\beq
ds^2 = a^2 \left[(1+2\Psi) d\tau^2 - (1+2\Phi)dx^2\right],
\eeq
where $\tau$ is the conformal time, $x$ is the comoving coordinate, $a$ is the scale factor, and $\Psi$ and $\Phi$ are 
the metric perturbations. We neglect neutrino anisotropic stress and set $\Phi = -\Psi$; this is a good 
approximation for modes that entered the horizon before $T \gtrsim \MeV$, since Weak interactions are still in equilibrium, so neutrinos behave as a perfect fluid. 
The equation of motion for the linear ALP field perturbation $a_1$ is~\cite{Hu:2004xd}
\beq
\ddot a_1 + 2\Hcon \dot a_1 + (k^2 + m_a^2 a^2)a_1 - (\dot \Psi -3\dot \Phi) \dot a_0 + 2 a^2 m_a^2 \Psi a_0 = 0,
\label{eq:alp_eom_perturbed}
\eeq
where the dots denote derivatives with respect to conformal time, 
$\Hcon = a H$ is the comoving Hubble parameter and $a_0$ is the background solution of Eq.~\ref{eq:alp_eom}. 
The oscillation-averaged equations for the 
energy density and velocity perturbations are obtained by
constructing approximate solutions in the $\Hcon/(m_a a)$ expansion as in Ref.~\cite{Hwang:2009js}.

We find the Fourier-space density $\delta_i$ and velocity divergence $\theta_i = i \vec{k}\cdot \vec{v}_i$ perturbation equations for $i=\phi$ (the EMD field),
$a$ (the ALP), and $r$ (SM radiation) to be:
\begin{subequations}
\begin{align}
  \dot \delta_\phi + \theta_\phi + 3\dot\Phi & = -a \Gamma_\phi \Psi,\\
  \dot \theta_\phi + \Hcon \theta_\phi - k^2\Psi & = 0,\\
  \dot \delta_r + \frac{4}{3}\theta_r + 4\dot\Phi & = a \Gamma_\phi \frac{\rho_\phi}{\rho_r}
  \left[\delta_\phi -\delta_r + \Psi\right],\\
  \dot \theta_r -\frac{1}{4}k^2 \delta_r - k^2\Psi & = a \Gamma_\phi \frac{\rho_\phi}{\rho_r}
  \left[\frac{3}{4}\theta_\phi -\theta_r\right],\\
  \dot \delta_a + \theta_a + 3\dot\Phi & =  -3\cnad^2\Hcon  \delta_a - 9\cnad^2 \Hcon^2 \theta_a/k^2 \label{eq:delta_a_avg}\\ 
  \dot \theta_a + \Hcon \theta_a - k^2\Psi & = +3\cnad^2 \Hcon \theta_a + k^2\cnad^2\delta_a \label{eq:theta_a_avg}\\
 k^2\Phi -3\Hcon\left[\Hcon\Psi-\dot\Phi\right] &= 4\pi G a^2\left[\rho_\phi\dels+\rho_r\delta_r+\rho_a\delta_a\right], \label{eq:grav_pot_avg}
\end{align}
\label{eq:perturbed_boltzmann_system}%
\end{subequations}
where the non-adiabatic sound speed $\cnad$ is 
\beq
\cnad^2 = \frac{k^2}{k^2 + 4 m_a^2 a^2}.
\label{eq:cnad_avg}
\eeq

We take standard adiabatic initial conditions for all fluids. These are discussed in detail in App.~\ref{app:boltzmann}. 
Note that this assumption for the ALP is not trivial. Under standard assumptions and high-scale 
inflation, quantum fluctuations of the ALP will generate isocurvature initial conditions at large scales. 
Consistency with the CMB then implies either a low scale of inflation or non-trivial inflationary dynamics, as discussed in Sec.~\ref{sec:relic_abundance}. For example, if the scale of inflation is low, the isocurvature component in the ALP is tiny. The adiabatic component vanishes until the ALP begins to oscillate. After oscillations begin, superhorizon modes of the ALP develop adiabatic perturbations, as we show 
in Appendix~\ref{sec:ic_from_eom}.

The system of equations~(\ref{eq:perturbed_boltzmann_system}) is nearly identical 
to those derived in Ref.~\cite{Erickcek:2011us} for the WIMP, except for the appearance of the 
sound speed terms, the initial conditions, and the generation of the relic abundance through misalignment.
As a result, modes that enter the horizon after oscillations and for which the sound speed is not important evolve as described in Ref.~\cite{Erickcek:2011us}. Likewise, the other fluids evolve as in Ref.~\cite{Erickcek:2011us}.

Before  numerically analyzing of Eq.~(\ref{eq:perturbed_boltzmann_system}), it is useful to study the 
behavior of the ALP density perturbations analytically. First, let us focus on longer wavelength modes, which enter the horizon after the ALP begins to oscillate, and for which sound speed 
effects can be neglected. The 
growing mode of perturbations that enter the horizon before RH is 
\beq
\delta_a(a,k) = 2\Phi_0 + \frac{2}{3}\left(\frac{k}{H_1}\right)^2 a \Phi_0 \;\;\; (a < \arh,\; k > \krh),
\label{eq:emd_alp_approx_sol}
\eeq
where $\Phi_0$ is the initial value of the gravitational potential (prior to reheating),
and $\krh$ is the comoving scale corresponding to the comoving horizon at reheating:
\beq
\krh = \Hcon(\arh) \sim \arh^{-1/2}.
    \label{eq:krh_def}
\eeq
This wavenumber can be expressed in physical variables~\cite{Erickcek:2011us}:
\beq
  \frac{\krh}{\keq}  = \frac{\krh}{\aeq H(\Teq)} \approx 5.9\times 10^6 \left(\frac{\TRH}{5\;\MeV}\right)\left(\frac{g_*(\TRH)}{10.75}\right)^{1/6},
\label{eq:krh_over_keq}
\eeq
where we used $\aeq = 1/(1+\zeq)$, $\zeq = 3365$ and $\Teq = 0.79\;\eV$~\cite{Ade:2015xua}.

After RH, these modes continue to grow logarithmically; assuming horizon entry occurs well before instantaneous reheating at $a=\arh$ and 
matching solutions in the two regimes, one finds that for $a > \arh$
\beq
\delta_a(a,k) = \frac{2}{3}\left(\frac{k}{\krh}\right)^2 \Phi_0  \ln\frac{e a}{\arh}\;\;\; (a > \arh, k > \krh).
\label{eq:approx_sol_after_rh_enhanced}
\eeq
Here the prefactor $\propto (k/\krh)^2$ encodes the period of enhanced growth between horizon entry and 
reheating; since ${\cal H}\sim a^{-1/2}$ during EMD, linear growth corresponds to
\begin{align}
\arh/\ahor= (k/\krh)^2.
\end{align} Modes that enter the horizon after RH evolve as in $\lcdm$; the solution is well-approximated 
by the fitting formula~\cite{Hu:1995en}
\beq
\delta(a,k) = \frac{10}{9} A \Phi_0 \ln \left(\frac{B a}{\ahor}\right) \;\;\; (a > \arh, k < \krh) 
\label{eq:approx_sol_rd}
\eeq
after horizon entry $(a \gg \ahor)$, with $A \approx 9.11$ and $B \approx 0.594$; 
the prefactor $10/9$ accounts for the transition from EMD to RD.

\begin{figure}
  \centering
  \includegraphics[width=0.47\textwidth]{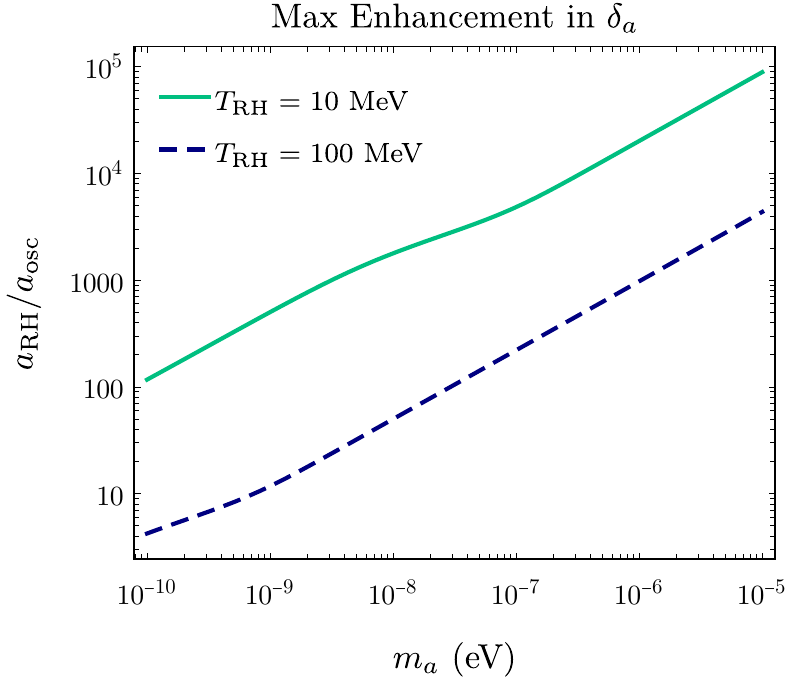}
  \caption{The ratio of scale factors at the onset of ALP oscillation and at reheating $a_{\rm{osc}}/a_{\rm{RH}}$, corresponding to the maximum possible enhancement in the growth of the dark matter perturbation $\delta_a$, for reheating temperatures $T_{\rm{RH}}=10$ and 100~MeV.
\label{fig:enhancement}}
\end{figure}

How much growth can an axion perturbation  undergo during EMD? Linear growth in the scale factor corresponds to $\delta_a(\TRH)/\delta_a(\Tosc) \sim (\kosc/\krh)^2$ (see Eq.~\ref{eq:approx_sol_after_rh_enhanced}), where $\kosc$ 
is the comoving horizon size when oscillations begin:
\begin{align}
\kosc  = m_a \aosc. 
\label{eq:kosc}
\end{align}
Therefore
\begin{align}
   \left( \frac{\kosc}{\krh} \right)^2&=\frac{\arh}{\aosc}   \approx \left(\frac{m_a}{H(\TRH)}\right)^{2/3} \nonumber\\
  & \sim 10^6 \left(\frac{m_a}{10^{-5}\;\eV}\right)^{2/3} \left(\frac{5\;\MeV}{\TRH}\right)^{4/3} \left(\frac{10.75}{g_*(\TRH)}\right)^{1/3}.
  \label{eq:aosc_over_arh}
\end{align}
Note also that during RD the temperature falls as $T\propto a^{-1}$, but during EMD, entropy release from the $\phi$ decay causes the temperature to fall as a smaller power,  $T\propto a^{-3/8}$~\cite{Scherrer:1984fd}. The fact that axion miniclusters are produced in an EMD cosmology relies both on this fact and on the linear growth of perturbations during EMD.

The only modes that could possibly grow by the amount~(\ref{eq:aosc_over_arh}) must already be inside the horizon at $\Tosc$, 
as well as being non-relativistic and below the Jeans scale (discussed below). 
This is an estimate that represents the maximum theoretically possible growth of a mode during EMD. 
We show this result in Fig.~\ref{fig:enhancement} for $\TRH=10$ and 100~MeV. We see that we can indeed achieve a duration of matter domination with $a_{\rm RH}\sim \mathcal{O}(10^4)$ after ALP oscillation (taking $a_{\rm osc}=1$), but that this requires $m_a \gtrsim 10^{-6}$~eV and a low reheating temperature. For low masses, the enhancement is quite small, since the ALP starts to oscillate later and so does not benefit as much from EMD.

The previous discussion neglected the effective ALP pressure. This approximation turns out to be excellent for modes that enter the horizon well after oscillation and well before reheating, which will be the most important for the formation of miniclusters. To see this, and to understand the effect of the sound speed on other modes, we combine Eqs.~(\ref{eq:delta_a_avg}) and (\ref{eq:theta_a_avg}) 
into a single second-order equation.
The exact result is complicated, 
but if we take $\Hcon \sim k \ll m a$ (i.e. $\cnad \ll 1$) then it can be expanded in these small quantities, 
with the result
\beq
\ddot \delta_a + \Hcon \dot \delta_a + \cnad^2 k^2 \delta_a = -k^2 \Psi - 3 \Hcon \dot \Phi - 3 \ddot \Phi .
\label{eq:2nd_order_delta}
\eeq
This equation was also obtained in Ref.~\cite{Arvanitaki:2009fg}. It is clear that the sound 
speed term competes with the gravitational driving term, and it prevents growth of the perturbations with $k$ greater than some Jeans scale $k_J$, which we estimate as follows.
First we analyze the EMD era, where the gravitational potential is dominated by 
    $\phi$ and is constant in time. When the ALP oscillations begin, the ALP density 
    contrast is $\delta_a \approx 2 \Phi_0$ (see Sec.~\ref{sec:pert_and_ic} and App.~\ref{app:boltzmann}). The sound 
    speed term prevents growth until it becomes comparable to the gravitational source term $k^2\Psi = -k^2\Phi$, 
    which occurs when $ k < m_a a$. Thus during EMD, $k_J \sim m_a a \equiv \kcompt$.\footnote{
    Note that this is a different scaling with scale factor and ALP mass than found in, e.g., Ref.~\cite{Hu:2000ke,Hlozek:2014lca,Arvanitaki:2009fg,Fairbairn:2017sil}, 
    due to the fact that the gravitational potential is dominated by the EMD field $\phi$, rather than the ALP itself.} 
Moreover, modes that enter the horizon during EMD but well after oscillations satisfy $k\ll \aosc m_a \ll a m_a \sim k_J$.
    On the other hand, since $k_J \sim \kcompt$, modes with $k>k_J$ during EMD are not well-described using the 
    oscillation-averaged equations used here. It is still true, however, that during this period 
    the density perturbation does not grow beyond the CDM adiabatic initial condition $2\Phi_0$, 
    as illustrated in the right panel of Fig.~\ref{fig:comoving_scales}.
    
    After reheating, the gravitational potential rapidly decays and oscillates, averaging to zero over cosmological time scales for 
    $k \gg \Hcon$. In this limit, the source terms on the RHS of Eq.~\ref{eq:2nd_order_delta} can be dropped, leading to an approximate solution
    \beq
    \delta_a \sim \sin\left(\frac{k^2}{2m_a\sqrt{\arh}} \ln \frac{a}{\arh} \right).
    \label{eq:sol_during_rd}
    \eeq
    The perturbation grows logarithmically as for a CDM fluid when the argument of the sine is small, and 
    begins to oscillate (with a constant 
    frequency in $\ln a$) when $k^2 \sim  2 m_a \sqrt{\arh}/\ln(a/\arh)$. 
    Combining the EMD and RD regimes 
    we find that the Jeans scale (expressed in physical units) is
    \beq
    k_J = \begin{cases}
      m_a a & (a \leq \arh) \\
      \frac{\arh \sqrt{2 m_a H(\arh)}}{\sqrt{\ln(a/\arh)}} & (a > \arh).
    \end{cases}
    \label{eq:jeans_def}
    \eeq


\subsection{Numerical Solutions}
\label{sec:numsol}

\begin{figure*}
  \centering
  \includegraphics[width=0.47\textwidth]{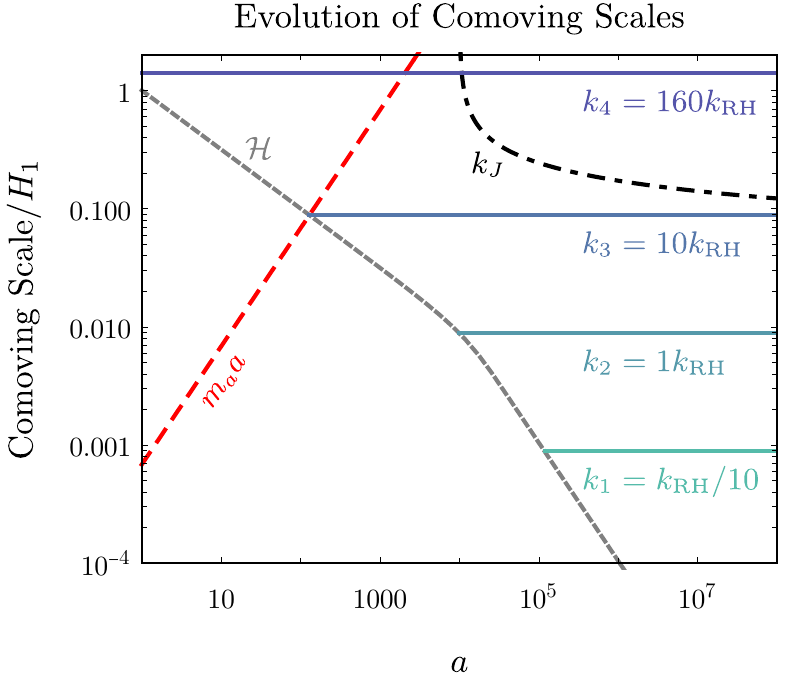}
  \includegraphics[width=0.47\textwidth]{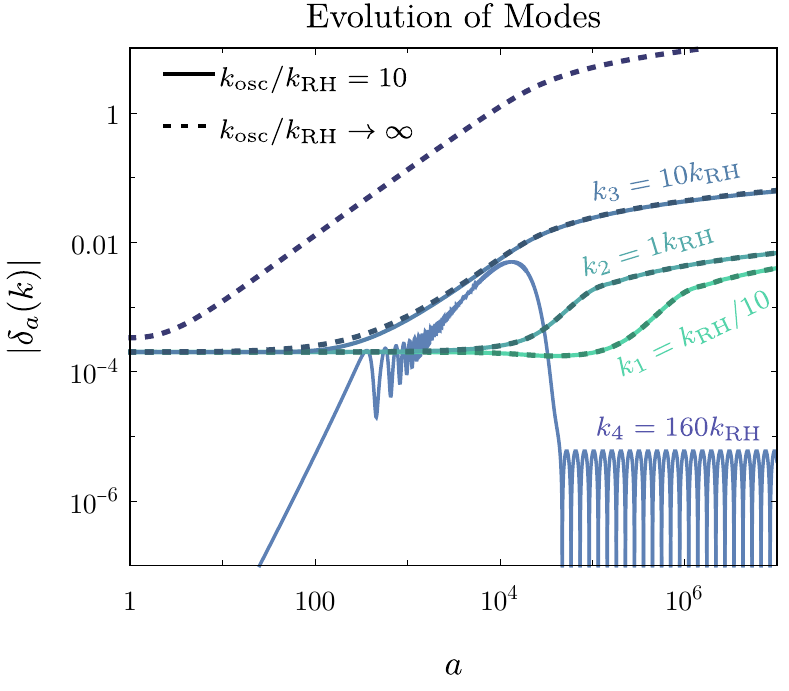}
  \caption{(Left) Comoving wavevectors $k_i$ of representative ALP perturbations (solid lines) compared to the evolution of comoving Hubble distance $\Hcon$ (short dashed gray), ALP mass $m_a$ (dashed red), and the Jeans scale $k_J$ (dot-dashed black). Each mode begins evolving when $k\approx \Hcon$ and its subsequent 
    behavior depends on its size relative to the other scales at a given time.
    The $\phi$ decay rate is such that reheating happens at $\arh \approx 10^4$ and the ALP mass is chosen such that $\kosc/\krh = 10$, 
    i.e. ALP oscillations begin before reheating. $k_1$ enters the horizon after reheating and evolves as in the standard cosmology; 
    $k_2$ enters at reheating and corresponds to physical scales much larger than the ALP Jeans scale; $k_3$ and $k_4$ 
    enter deep during the EMD; $k_3$ ($k_4$) is physically larger (smaller) than the Jeans scale.
    (Right) Evolution of modes with different wavevectors $k$ (same as in the left panel)
    as a function of the scale factor (solid lines) compared to the CDM case (dotted lines), corresponding to $\kosc/\krh \rightarrow \infty$ or, equivalently, 
    large $m_a$. The growth of modes with $k > k_J \sim \kosc$ is suppressed by the effective ALP pressure, while those with $k < k_J$ are identical to the CDM case. 
    Modes that enter the horizon before RH experience enhanced growth due to EMD.
\label{fig:comoving_scales}}
\end{figure*}

In this section we solve the full linearized system of equations described above  numerically and study the growth of ALP 
perturbations. We interpret the evolution of perturbations with different $k$ in terms of their relation to the key physical scales:
\begin{itemize}
  \item the horizon size $\Hcon$
  \item the Compton scale $\kcompt = m_a a$
  \item the horizon size at RH, $\krh$, defined in Eq.~\ref{eq:krh_def} 
  \item the horizon size when ALP oscillations begin, $\kosc$, defined in Eq.~\ref{eq:kosc}
  \item the Jeans scale, $k_J$, defined in Eq.~\ref{eq:jeans_def}, which encodes the effects of the ALP effective pressure.
\end{itemize}
In Fig.~\ref{fig:comoving_scales} we illustrate the evolution for several values of $k$; the left panel shows the relation of 
these wavenumbers to the key scales above. The right panel shows the time evolution of the corresponding perturbations.  
The mode $k_1 < \krh$ enters during RD, and is always smaller 
than the Jeans and Compton scales, so its evolution is CDM-like with no 
enhancement from EMD or suppression from the ALP effective pressure.\footnote{There is a brief kick of non-logarithmic growth as modes cross the horizon due to gravitational driving~\cite{Hu:1995en}. This is particularly clear for the $k_1$ mode and it is unrelated to the period of EMD.}

The mode $ k_2 = \krh$ entered at reheating, while the Jeans scale is unimportant, so it evolves as a CDM perturbation.
The mode $k_3 > \krh$ enters during EMD, while 
quasi-relativistic, and it is slightly sensitive to the Jeans scale at the start of its evolution. This can be seen in the right-hand plot in the small suppression in the growth of the $k_3$ mode relative to the same mode in the CDM case as they both enter the horizon.
Finally, the mode  $k_4 $ enters the horizon while relativistic.
After it becomes 
non-relativistic however ($a = 5000$), we see that its amplitude is suppressed and it undergoes 
rapid oscillations driven by gravity and ALP effective pressure.
We note that ALP oscillation-averaged equations (Eqs.~\ref{eq:perturbed_boltzmann_system}) are not adequate in this case before the onset of oscillations; we therefore 
solve for its early evolution using the field equation~\ref{eq:alp_eom_perturbed}. 
Its late-time evolution matches onto the solutions of the oscillation-averaged equations with 
adiabatic initial conditions discussed above. We therefore only use these equations in the remainder of this work.

\begin{figure*}
  \centering
  \includegraphics[scale=0.7]{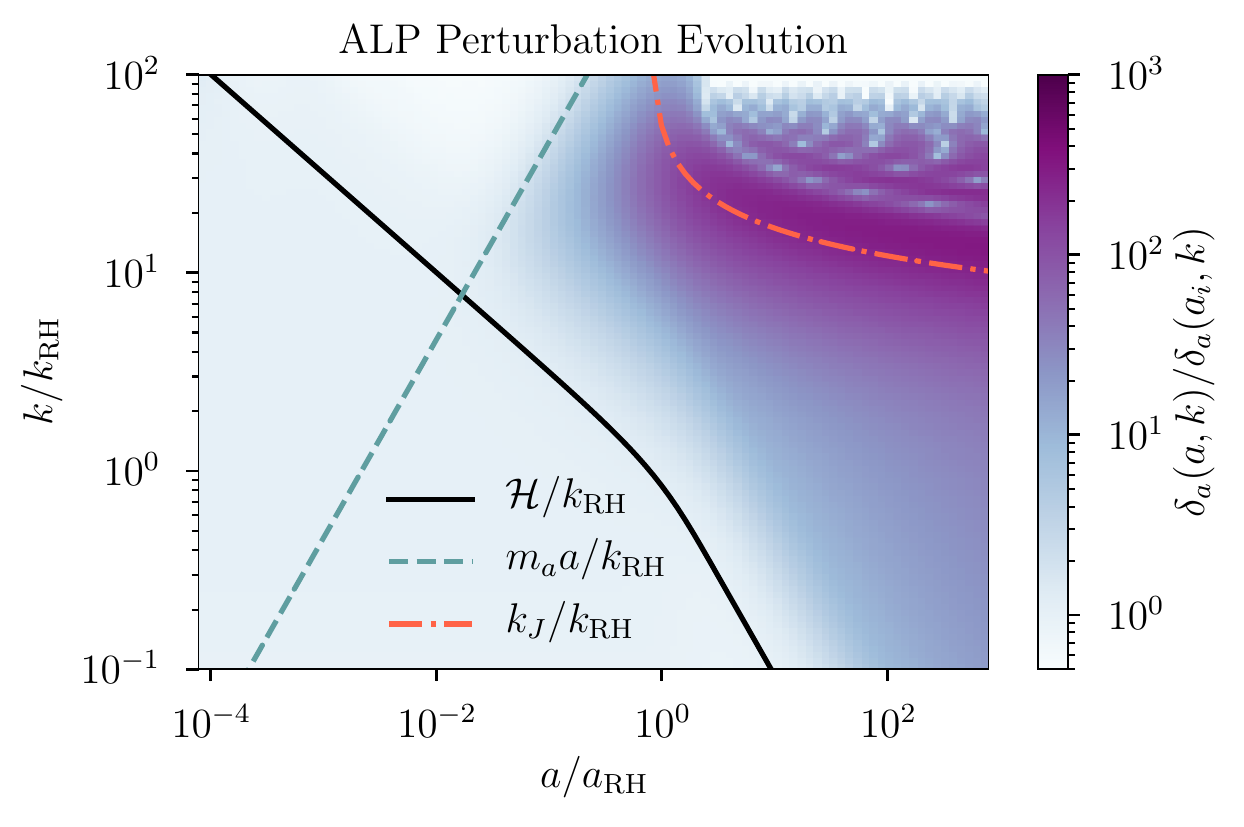}
  \includegraphics[scale=0.7]{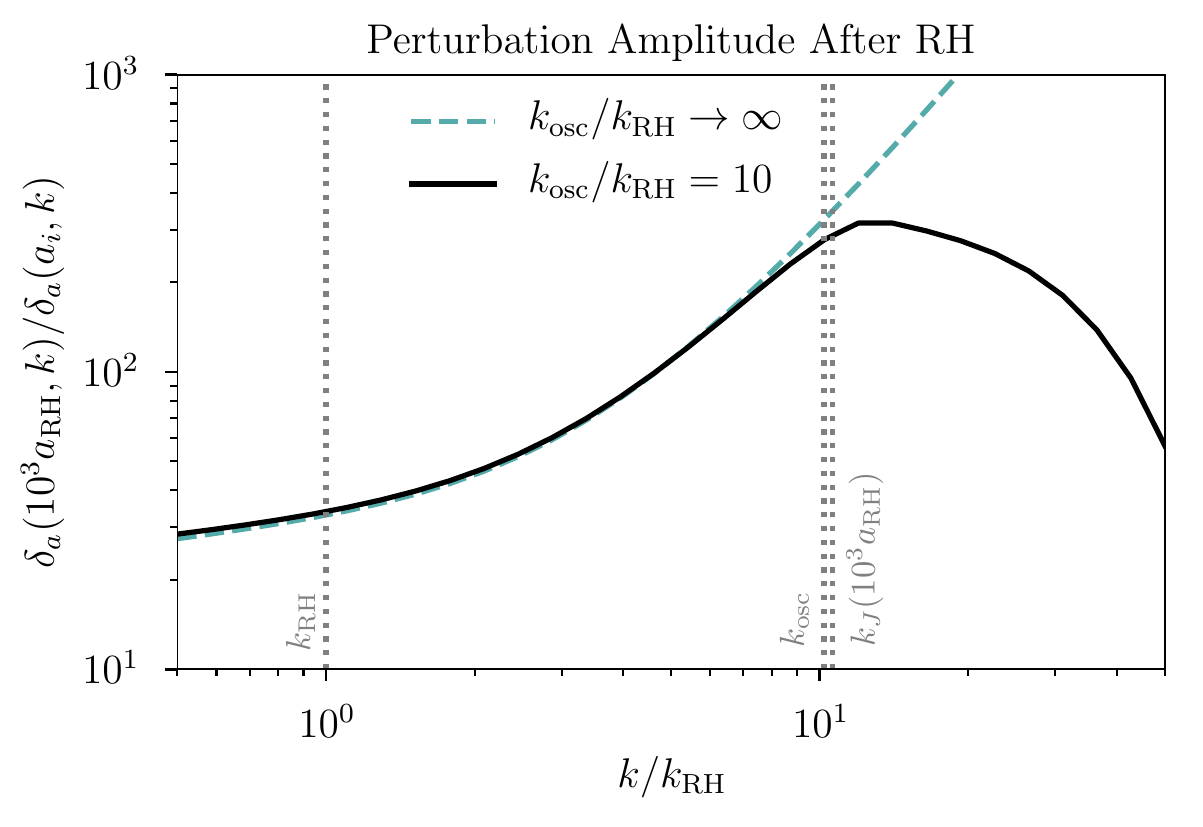}
  \caption{Growth of ALP perturbations $\delta_a$ with different comoving wavenumbers. 
    The left panel shows the evolution of $\delta_a$ (normalized to its initial value) 
    in the plane of $a/\arh$ and $k/\krh$. Important comoving scales are 
    indicated by lines (in analogy to Fig.~\ref{fig:comoving_scales}): 
    the comoving Hubble horizon $\Hcon$ (solid black), the comoving Compton scale $m_a a$ (teal dashed) and 
    the Jeans scale (red dot-dashed).  
    The right panel shows the amplitude of ALP perturbations well after reheating (at $a \sim 10^3 \arh$) with and without 
  the ALP sound-speed effects, solid and dashed lines respectively. Key scales are shown by vertical dashed gray lines. 
  In both panels, the ALP mass is chosen such that $\kosc/\krh = 10$.
  The growth of modes with $k \gtrsim k_J\sim \kosc$ is suppressed because of the effective 
  ALP pressure. These modes also enter the horizon prior to ALP oscillations, when the ALP field still behaves like a cosmological constant. 
  Modes with $k/\krh \lesssim 1$ enter after reheating and therefore evolve as in 
  standard cosmology. Note that the $k > k_J$ part of the right panel is the envelope of the oscillating function 
  shown in the left panel. 
 }
  \label{fig:mode_amplitude}
\end{figure*}

The previous discussion can be used to understand Fig.~\ref{fig:mode_amplitude}, which shows the evolution of a wide range of $k$ modes in the left panel; 
the right panel illustrates the suppression of growth due to the effective pressure well after reheating at $a=10^3\arh$.
The most important features are evident in the left panel: modes that enter the 
horizon benefit from EMD, while those with $k\gg k_J \sim \kosc$ are suppressed. 
The magnitude of the suppression is scale and somewhat time-dependent (due to the scale factor dependence of $k_J$ -- see Eq.~\ref{eq:jeans_def}).
In the following sub-section we combine these numerical solutions with late-time 
growth of perturbations during standard radiation and matter domination. The goal is 
to evaluate the ALP perturbation power spectrum and smoothed density variance at 
various redshifts. In order to simplify these considerations we will model the small-scale 
suppression of power as a sharp cutoff at $k = \kosc$.
This is an approximation as the actual fall-off is much smoother -- see the right panel of Fig.~\ref{fig:mode_amplitude}. 
However, this simplification will enable a fast exploration of the minihalo parameter space without having to solve 
the full Boltzmann system for each $(m_a, \TRH)$. A more sophisticated treatment of these small-scale effects would 
involve using the analytic solution during RD (Eq.~\ref{eq:sol_during_rd}) to evaluate the density 
contrast and the corresponding transfer function at late times. This approximation is 
appropriate for the gravitational signals we will study in Sec.~\ref{sec:analysis}, since 
they are sensitive to the largest surviving minihalos which originate as $k<\kosc$ perturbations 
in models with $\krh \ll \kosc$ (i.e. when there is an extended period between the start of ALP oscillations and 
end of reheating). In scenarios where there is only a mild hierarchy $\krh\lesssim \kosc$, the sharp 
cut-off underestimates the abundance of minihalos and their survival probability, and therefore leads to conservative 
estimates of signal rates.

\subsection{Growth and Transfer Function}
The minicluster abundance depends on the evolution of ALP 
perturbations towards collapse. For our choice of adiabatic initial conditions this happens at or 
after the standard matter-radiation equality. In order to match onto this ``standard'' growth period 
we modify the semi-analytic prescription of Ref.~\cite{Erickcek:2011us} to account for ALP mass effects. 
The quantity of interest is the time-dependent fluctuation variance smoothed over comoving length
scales $R$
\begin{align}
\sigma^2(a,R) & = \int \frac{d^3 k}{(2\pi)^3} \langle \delta_a(a,k)^2 \rangle W(kR)^2\nonumber \\ 
& = \frac{1}{2\pi^2}\int dk k^2 W(k R)^2 |TD(a,k)|^2 P(k) \, .
\label{eq:PS_sigma_squared}
\end{align}
The averaging is achieved through the use of a window function $W(kR)$, which we take to be a spherical top-hat in real space. In the second line we wrote the time-dependent ALP perturbation as
\beq
\delta_a(a,k) \approx \frac{2k^2}{5\Omega_m H_0^2} TD(a,k) \mathcal{R}(k) \;\;\; (a > \aeq),
\label{eq:delta_parametrization}
\eeq
where $\mathcal{R}$ is the primordial scalar curvature fluctuation amplitude that determines the 
initial conditions for the evolution of $\delta_a$; this equation defines $TD(a,k)$, the scale-dependent growth function which we discuss in more detail below. 
The primordial matter power spectrum $P(k)$ is related to the power spectrum of $\mathcal{R}$ by 
\beq
P(k) = \left(\frac{2 k^2}{5\Omega_m H_0^2}\right)^2 P_\mathcal{R} (k),
\label{eq:primordial}
\eeq
where 
\beq
P_\mathcal{R}(k) = \frac{2\pi^2}{k^3}A_s\left(\frac{k}{k_0}\right)^{n_s-1},
\label{eq:primordial_curvature_ps}
\eeq
is set by inflation, with $k_0 = 0.05\;\Mpc^{-1}$ and Planck best-fit values of $\ln(10^{10} A_s) =3.044\pm 0.014$ and $n_s =0.965 \pm 0.004$ ~\cite{Aghanim:2018eyx}.
We note that the scales we are interested 
are far smaller than those probed by the CMB; thus the assumption of a power-law spectrum with a constant $n_s$ amounts to a 
significant extrapolation. While the uncertainties in the $n_s$ measurement do not qualitatively affect the results below, 
we emphasize that the $k\gg k_0$ part of the matter power spectrum we are studying has not been measured directly.

The scale-dependent growth function $TD(a,k)$ contains both the dynamics of the Boltzmann equations solved in the previous section and the post-reheating growth of ALP overdensities. For cosmologies without baryons or the cosmological constant, $TD$ factorizes as 
$TD(a,k) = T(k) D(a)$, where $T$ and $D$ are the standard transfer and growth functions, respectively. 
The normalization factors in Eq.~(\ref{eq:delta_parametrization}) are chosen such that 
$T(k) \approx 1$ for modes that entered the horizon after matter-radiation equality and $D(a) \approx a$ deep 
in the matter domination era. The factor of $k^2$ in Eq.~(\ref{eq:delta_parametrization}) combines with $D(a)=a$ to become $k^2a=a/\ahor$, representing linear growth during the most recent era of matter domination. The transitions from EMD to RD at $\TRH$, and from RD to MD at $\Teq$ introduce two characteristic scales $\keq$ and $\krh$ into the transfer function. Schematically, $T$ behaves as
\beq
T(k) \sim \begin{cases}
  1 & k < \keq \\
  (k/\keq)^{-2} \ln (k/\keq)& \krh \geq k \geq \keq \\
  (\krh/\keq)^{-2} & \kosc > k > \krh \\
  0 & k > \kosc
\end{cases},
\label{eq:approx_transfer_scaling}
\eeq
The first line reflects the fact that modes with $k<\keq$ enter the horizon after matter-radiation equality and their growth 
is entirely captured by $D(a)$. The second line applies to modes that enter the horizon during radiation domination (after reheating and before MRE); for these modes, the transfer function measures the deviation from linear growth. The factor of $(k/\keq)^{-2}$ removes this linear growth from $k^2 D$ and the factor of $\ln (k/\keq)$ restores the standard logarithmic growth during RD. The transfer function plateaus for modes that enter the horizon before reheating, simply removing linear growth between reheating and equality, but preserving it at higher scales. Finally, modes 
that enter before oscillations begin, are suppressed due to the ALP effective pressure (note that this sharp cut-off is only a rough approximation -- 
the actual fall-off is $k$ and $a$-dependent as discussed in the previous section).

In a universe with a non-negligible baryon abundance and a late-time period 
of dark energy domination, $TD$ does not factorize. The former ensures that the growth of 
modes below the baryonic Jeans scale is slower (since baryons are pressure-supported on 
these scales), while the latter affects the late-time evolution of structure. 
In both cases the growth rate becomes scale-dependent as can be seen by 
solving the Meszaros equation~\cite{Erickcek:2011us}.
However, the 
 approximate scaling relations in Eq.~(\ref{eq:approx_transfer_scaling}) can still be used to qualitatively understand our results. The transfer function can also be altered through the presence of a non-zero velocity dispersion for dark matter, which washes out structure at low masses. ALPs produced through the misalignment mechanism have a momentum dispersion which is of order the Hubble scale at the time of oscillation, which is negligible~\cite{Visinelli:2009kt}.

\begin{figure}
  \centering
  \includegraphics[width=0.47\textwidth]{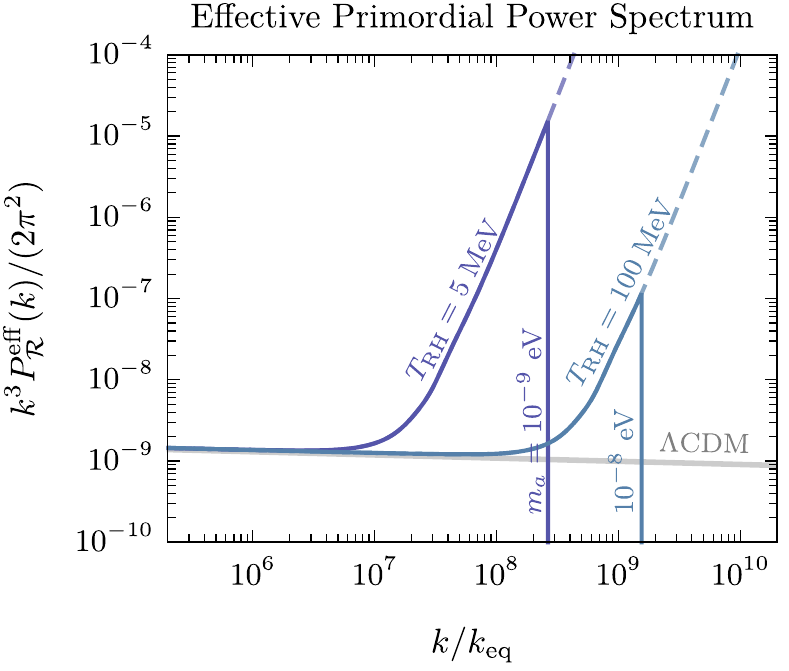}
  \caption{The dimensionless effective primordial curvature power-spectrum $ k^3 P_{\mathcal{R}}^\mathrm{eff}(k)/(2\pi^2)$ that encodes EMD structure growth 
    at small scales. The grey band corresponds to the power spectrum in the standard cosmology as in Eq.~\ref{eq:primordial_curvature_ps}. 
    The solid blue lines correspond to $\TRH=5$~MeV, $m_a = 10^{-9}$ eV (upper) and $\TRH = 100\;\MeV$, $m_a = 10^{-8}$ eV (lower).  The sharp cut-offs at small scales in the small $m_a$ cases roughly approximate 
    the suppression of power for modes with $k>k_J\sim\kosc$ due to the effective ALP sound speed. This suppression is absent in the CDM case (dashed lines), 
    corresponding to $m_a \rightarrow \infty$.
\label{fig:power_spectrum}}
\end{figure}

In order to 
facilitate the exploration of parameter space, we follow Ref.~\cite{Erickcek:2011us} in 
defining a semi-analytic approximation to $\delta_a(a,k)$ that allows us to propagate our 
results into standard matter domination. The linearity of the Boltzmann equations and 
decoupling of different $k$-modes allows us to approximate the full scale-dependent 
growth function as 
\beq
TD(a,k) = TD_{\lcdm}(a,k) R(k),
\label{eq:sdgf_parametrization}
\eeq
where
\beq
R(k) = \frac{\delta_a (a,k)}{\delta_a^{(\lcdm)}(a,k)} 
     \sim \begin{cases}
       1 & k < \krh \\
       \frac{(k/\krh)^{2} }{\ln(k^2/\krh^2)} & \kosc \geq k \geq \krh \\
     0 & k > \kosc
   \end{cases}.
   \label{eq:R_def_and_scaling}
\eeq
The growth of perturbations after equality cancels in the ratio such that $R$ does not 
depend on $a$. 
This parametrization is useful because $R(k)$ encodes the effects of EMD and the ALP mass relative to $\lcdm$, while $TD_{\lcdm}(a,k)$
can be computed  using a standard Boltzmann code~\cite{Lewis:1999bs,Blas:2011rf} or using well-known fitting 
formulae~\cite{Hu:1995en}.
The precise form of $R(k)$ and the computation of $TD_{\lcdm}$ are described in Appendix~\ref{sec:fitting_function}.
The scalings in Eq.~(\ref{eq:R_def_and_scaling}) follow from the approximate solutions in Eqs.~(\ref{eq:approx_sol_after_rh_enhanced}).
Since $TD_{\lcdm} \sim 1/k^2$ at $k \gg \keq$ and $R\sim k^2$ for 
modes that entered during EMD, we see that the rescaling by $R$  results in the ``flattening'' of the transfer function 
as indicated in the third line of Eq.~(\ref{eq:approx_transfer_scaling}).\footnote{To qualitatively summarize the slightly ridiculous parametrization $\delta\sim k^2 TD_{\Lambda CDM}R$: $k^2 D$ puts in linear growth for all modes for all times, $TD_{\Lambda CDM}$ removes all linear growth prior to matter-radiation equality for modes entering during this time, and $R$ restores it again between ALP oscillations and reheating for modes entering during this time, while setting to zero modes that enter the horizon before oscillations begin.}

We can capture the EMD-enhanced growth in terms of 
  an effective primordial power spectrum $P_\mathcal{R}^{\mathrm{eff}}(k) = P_\mathcal{R}(k)\times R(k)^2$ that can 
    be used as input to $N$-body simulations. This effective power spectrum is shown in Fig.~\ref{fig:power_spectrum} 
    as a function of $k/\keq$.
    The grey band is for the standard cosmology as in Eq.~\ref{eq:primordial}. The blue lines correspond to our scenario for two different choices of $\TRH=5$ and 100~MeV. Modes with $k<\krh$ enter the horizon after reheating and evolve as in the standard cosmology. The wavenumbers of those modes can be estimated using Eq.~\ref{eq:krh_over_keq}. For modes $\kosc \geq k \geq \krh$ the power spectrum grows as $k^{3+n_s}$ as in Eq.~\ref{eq:R_def_and_scaling}. At large $k$ power is cut off by the Jeans scale. The dashed blues lines correspond to the heavy limit $m_a \to \infty$, for which there is no Jeans scale cutoff, and the solid blue lines correspond to $m_a=10^{-9}$~eV and $10^{-8}\;\eV$.

\begin{figure}
  \centering
  \includegraphics[width=0.47\textwidth]{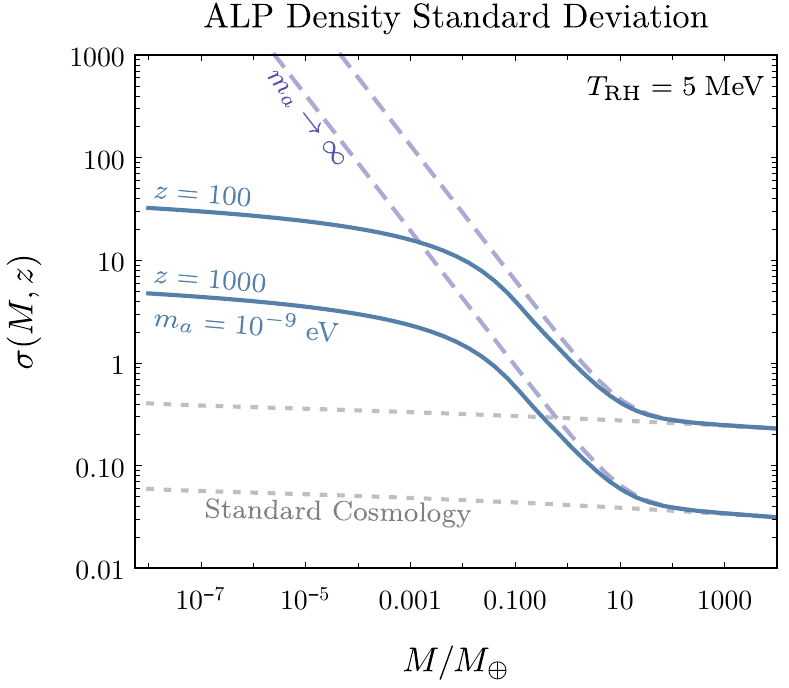}
  \caption{The density fluctuation standard deviation for a standard cosmology (dotted lines) and EMD cosmology with reheating at $\TRH=5$~MeV 
    with $m_a = 10^{-9}\;\eV$ and $m_a \rightarrow \infty$ (solid and dashed lines, respectively). For each model, 
    the lower and upper lines correspond to $z = 1000$ and $z = 100$.
\label{fig:density_variance}}
\end{figure}

In Fig.~\ref{fig:density_variance} we show the density standard deviation $\sigma$ for $\TRH = 5\;\MeV$, different choices of $m_a$ and 
two different times, $z=1000$ and $z=100$, illustrating the linear growth of density perturbations during standard 
matter domination. 
Its value in the standard cosmology is indicated by the dotted gray lines. In this case there is no enhancement at 
large $k$, corresponding to small $M$. The dashed lines show the small-scale enhancement in models with a large 
ALP mass where the effective pressure is not important. In this regime, 
\begin{align}
\sigma\propto \frac{1}{1+z}M^{-2/3}\krh^{-2}\propto\frac{1}{1+z}M^{-2/3}\TRH^{-2}
\end{align}
In contrast, if $m_a$ is small, the 
effective pressure introduces a small scale cut-off in the EMD mode ``amplification'', leading to a flattening 
of $\sigma$ at small $M$. Larger $\TRH$ reduce the largest mass scales that benefit from increased growth during EMD. 
In the following section, we use these results to estimate minihalo size, density and distribution.

\section{Minicluster Abundance}
\label{sec:halo_function}
Early matter domination leads to enhanced structure growth for scales that entered the horizon 
before reheating. Given the bound on $\TRH \gtrsim 5\;\MeV$, the \emph{maximum} mass of 
fluctuations that benefit from this enhancement simply corresponds to the matter mass enclosed 
by the horizon when $T = \TRH$:
\beq
\MRH = 250\MEarth\left(\frac{5\;\MeV}{\TRH}\right)^3\left(\frac{10.75}{g_*(\TRH)}\right)^{1/2}.
\label{eq:mrh_approx}
\eeq
Perturbations with $M > \MRH$ evolve as in standard cosmology, while those with $M<\MRH$ 
experience enhanced growth. However, perturbations in the ALP fluid only 
begin to grow like matter after the oscillations have begun, so there is also a \emph{minimum} 
mass of fluctuations, corresponding to the mass inside the horizon at the start of oscillations, even in standard cosmology without EMD:
\beq
\Mosc \approx 2.7\times 10^{-11}\msun \Omega_a \left(\frac{10^{-5}\;\eV}{m_a}\right)\left(\frac{5 \;\MeV}{\TRH}\right), 
\label{eq:mosc_approx}
\eeq
where we used Eqs.~(\ref{eq:aosc_over_arh}) and~(\ref{eq:krh_over_keq}) and $\Omega_a$ is given in Eq.~(\ref{eq:Omega_EMD}).
It is these objects that are usually called miniclusters, especially in models with post-inflationary PQ 
breaking where they collapse very early. The purpose of this section is to show 
that these smallest miniclusters are assembled into larger clumps, since EMD enhances the growth 
of density fluctuations over a range of scales.\footnote{Coincidentally, in the standard $\Lambda\rm{CDM}$ scenario where the dark matter is a thermal relic whose density is set through freeze-out, the smallest gravitationally bound structures are also approximately Earth-mass microhalos~\cite{Diemand:2005vz}. The formation of these structures is determined by the time of kinetic decoupling from the Standard Model thermal bath. This occurs when the temperature has dropped a further factor of 10-1000 after freeze-out~\cite{Bringmann:2009vf}. For a 100~GeV WIMP this corresponds to a temperature of order $10-100$~MeV, similar to $\TRH$ in our scenario.}

We estimate the statistics of collapsed ALP DM objects  as a function of size and redshift 
using the Press-Schechter (PS) formalism~\cite{Press:1973iz,Bond:1990iw} with the results from linear theory discussed in the previous sections. 
We wish to estimate the mass spectrum of miniclusters,  their sizes and densities, and their assembly history.

\subsection{Halo Function}

The central assumption of PS is that the fraction 
of mass in structures of size $\sim R$ is equal to the probability that the smoothed density contrast $\delta_R$ exceeds a threshold $\delta_c$. 
The critical density 
contrast $\delta_c$ can be estimated from spherical collapse, with the result that 
overdensities with $\delta_R = \delta_c = 1.686$ (as derived in linear perturbation theory) should have collapsed; 
this number is insensitive to the precise cosmological model, i.e. variations in $\Omega_\Lambda$ and $\Omega_m$, as long as the collapse 
occurs during matter domination~\cite{Galform:2010}.\footnote{Collapse can also occur during RD following the end of EMD. In this 
    regime the collapse criterion is different -- see, e.g., Ref.~\cite{Blanco:2019eij}. Objects forming during this era 
    would be very light and unimportant for the gravitational probes we consider in Sec.~\ref{sec:analysis}. We therefore 
focus on objects that reach non-linearity at or after standard matter-radiation equality.} 
Using this prescription, the fraction $f$ of matter in objects of mass in the range $[M,M+dM]$ at redshift $z$ is 
\beq
df(M,z) = 
\sqrt{\frac{2}{\pi}} \frac{\delta_c}{M\sigma} \left|\frac{d\ln \sigma}{d\ln M}\right| \exp\left(-\frac{\delta_c^2}{2\sigma^2}\right) dM.
\label{eq:ps_mass_fraction}
\eeq
The halo function is the number density of collapsed objects in this mass range:
\beq
n(M,z) = \frac{\rho}{M} \frac{df}{dM}.
\eeq
Equation~\ref{eq:ps_mass_fraction} can be integrated to find the fraction of matter contained in collapsed 
objects with a mass in the range $[M_1, M_2]$
\beq
F(M_1, M_2) = \erf\left(\frac{\delta_c}{\sqrt{2}\sigma(M_2,z)}\right) - \erf\left(\frac{\delta_c}{\sqrt{2}\sigma(M_1,z)}\right). 
\label{eq:fraction_in_range}
\eeq
\begin{figure*}
  \centering
  \includegraphics[width=0.47\textwidth]{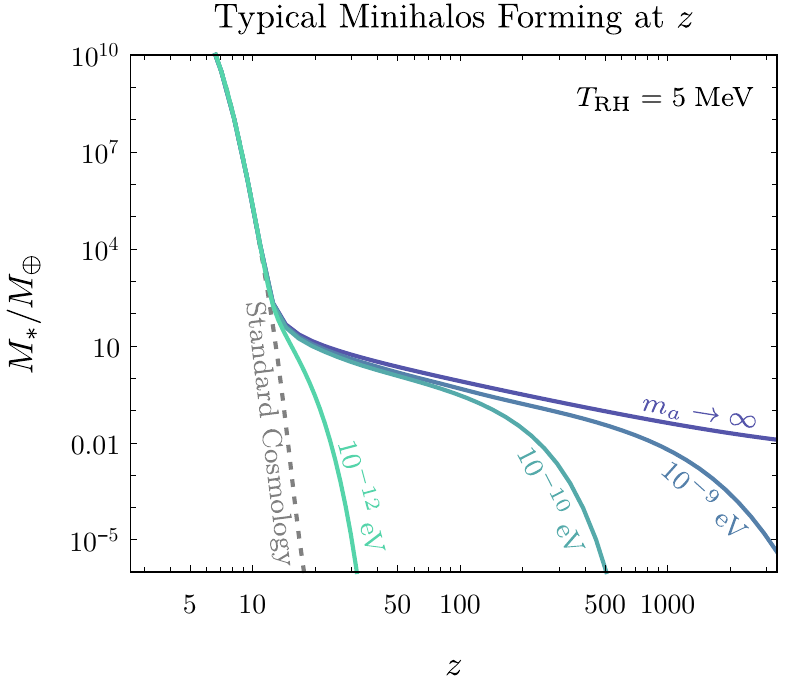}
  \includegraphics[width=0.47\textwidth]{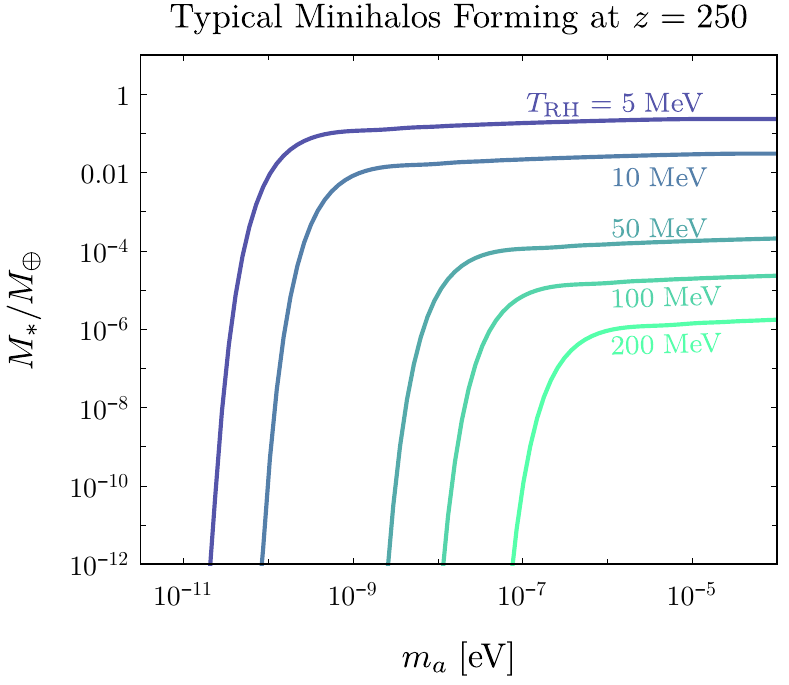}
  \caption{
    The typical mass $M_*$ (in units of the Earth mass) of minihalos forming at $z$, defined by $\sigma(M_*,z) = \delta_c$. 
    Results are shown as a function of redshift at fixed reheat temperature $\TRH = 5\;\MeV$
    for different ALP masses (left panel) and for fixed redshift $z=250$ as a function of the ALP 
    mass for different $\TRH$ (right panel). We expect that minihalos forming before this redshift will survive 
    tidal disruption through encounters with stars in the galaxy -- see Sec.~\ref{sec:minicluster_survival}.
    In the right panel the curves remain flat for higher values of $m_a$.
    \label{fig:mstar_plot}
  }
\end{figure*}
It is useful to define a characteristic mass $M_*(z)$,
\beq
\sigma(M_*,z) = \delta_c.
\label{eq:mstar_def}
\eeq
At a given redshift $z$, $M_*$ gives the mass of the typical structures 
that collapse at this time. The fact that $M_*$ is a monotonically increasing function of redshift 
suggests that structures are formed hierarchically in this model.
Using Eq.~(\ref{eq:PS_sigma_squared}) we see that 
$\sigma \sim M^{-(n_s+3)/6}$ for $M < \MRH$. Therefore 
we can approximately solve for $M_*(z)$:
\beq
\label{eq:mstar_estimate}
M_*(z) \approx \MRH \left(\frac{c}{1+z}\right)^{6/(n_s + 3)}
\eeq
where $c \sim 2.4-2.8$ for $\TRH \sim [5,100]\;\MeV$ is a slowly varying function of $\TRH$. 
This scaling is valid for $z \gtrsim 20$ and $m_a \rightarrow \infty$. 
In Fig.~\ref{fig:mstar_plot} we show the numerical solution for $M_*$ as a function of the 
collapse redshift in the left panel and the ALP mass in the right panel 
for several choices of the reheating temperature.
In the limit of a ``heavy'' ALP we find a result similar to that of 
Ref.~\cite{Erickcek:2011us} for CDM.  For small ALP masses, however, there is a noticeable suppression of 
minicluster formation. As explained in Sec.~\ref{sec:numsol}, this is due to the delayed onset 
of oscillations of the ALP, which reduces the amount of EMD linear growth experienced by the perturbations. 
Since this linear growth factor is $\sim \Tosc/\TRH$, the ``cutoff'' occurs at larger $m_a$ for larger reheat temperatures, as is evident in the right panel of Fig.~\ref{fig:mstar_plot}. The minimum $m_a$ needed to form minihalos at $z$ for a given $\TRH$ may be estimated by requiring the total amount of growth between oscillations and $z$ to be of order $10^5$,
\begin{align}
  10^5\approx  \left(\frac{\kosc}{\krh}\right)^2\log\left(\frac{\TRH}{T_{\rm eq}}\right)\left(\frac{1+z_{\rm eq}}{1+z}\right).
\end{align}
where the first factor is given by Eq.~(\ref{eq:aosc_over_arh}).
For thermal DM, an analogous cutoff in the power on small scales occurs in the presence of non-zero DM velocities and the resulting 
free-streaming~\cite{Erickcek:2011us}. The choice of $z=250$ in the right panel is motivated in Sec.~\ref{sec:minicluster_survival}: minihalos 
forming later than this are expected to undergo significant tidal disruption in stellar encounters.

The Press-Schechter formalism is based on the assumption of spherical collapse, and when compared with the results of numerical simulations overpredicts the amount of structure at the smallest scales and underpredicts the amount at larger scales~\cite{Jenkins:2000bv}. Better agreement with simulations in $\Lambda\rm{CDM}$ can be achieved using formalisms such as Sheth-Tormen~\cite{Sheth:1999su,Sheth:2001dp} which allow for ellipsoidal collapse. It would be interesting to study the implications of ellipsoidal collapse for ALP miniclusters in future work.

\begin{figure*}
  \centering
  \includegraphics[width=0.9\textwidth]{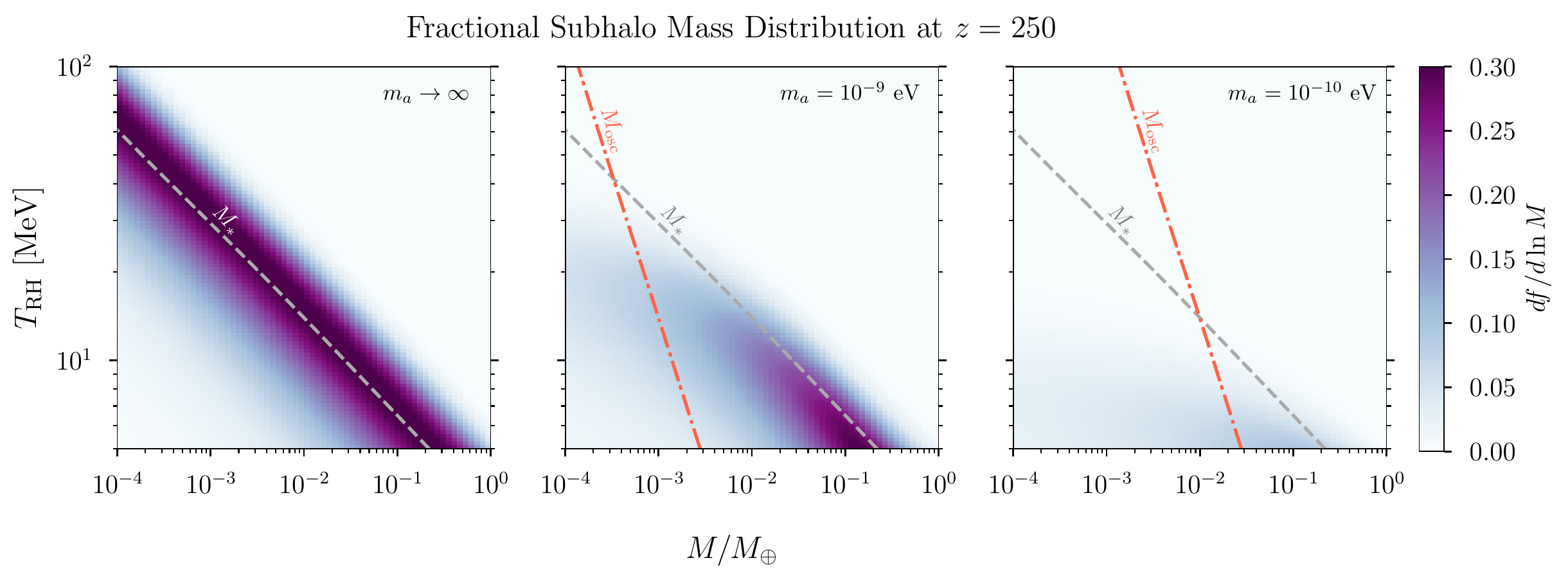}
  \caption{
    Fractional minicluster mass distribution $df/d\ln M$ at $z=250$ as a function of minicluster mass $M$ and reheating temperature $\TRH$ 
    for $m_a\to \infty$ (left panel), $m_a=10^{-9}$~eV (middle panel) and $m_a =10^{-10}$~eV (right panel). 
    Minicluster formation is suppressed for large $\TRH$ and small ALP masses.
    In each panel the gray dashed line shows the approximate value of $M_*$ (from Eq.~\ref{eq:mstar_estimate}), the mass of a typical minicluster forming at this 
  redshift in the CDM ($m_a\rightarrow \infty$) limit. The dot-dashed red line shows $\Mosc$, the DM mass within the horizon at the start of oscillations.
  Minihalos forming after $z=250$ are expected to undergo significant tidal disruption in encounters with stars -- see the discussion in Sec.~\ref{sec:minicluster_survival}.
    \label{fig:fractional_abundance_plot}
  }
\end{figure*}

In Fig.~\ref{fig:fractional_abundance_plot} we show the fractional minicluster abundance at a redshift of $z=250$ as function of the minicluster mass $M$ and $\TRH$. For finite ALP masses and at larger $\TRH$, there is less time during EMD for perturbations to grow, and the abundance peaks at smaller $M$. If the ALP is too light and $\TRH$ too high, oscillations begin after EMD ends, and the minicluster abundance is suppressed on all scales. 

\begin{figure}
  \centering
  \includegraphics[width=0.47\textwidth]{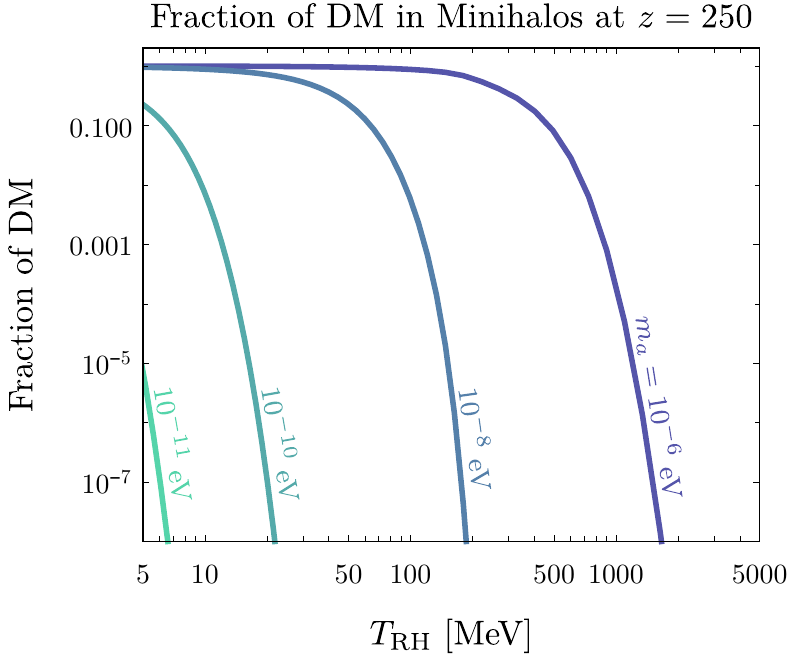}
  \caption{Fraction of ALP DM bound in minihalos of mass $M \in [\Mosc,\MRH]$ at $z=250$ as a function of $\TRH$ for different choices of the ALP mass. 
    In models with heavier ALPs and lower reheating temperatures, nearly all of DM is bound into minihalos at early times. 
    Estimates in Sec.~\ref{sec:minicluster_survival} indicate that minihalos forming at $z\gtrsim 250$ are resilient to 
    tidal disruption, suggesting that the fraction of ALPs in minihalos evaluated at early times is robust, i.e. 
    it remains roughly constant through the remaining evolution.
  \label{fig:fraction_in_minihalos}}
\end{figure}

Another quantity of interest is the fraction of DM mass in minihalos, which is given in Eq.~\ref{eq:fraction_in_range} and 
  corresponds to the area under the constant-$\TRH$ slices of the distributions in Fig.~\ref{fig:fractional_abundance_plot}. 
  This equation should be evaluated at a high enough redshift that standard $\Lambda$CDM-like halos have not 
  started to form. In Fig.~\ref{fig:fraction_in_minihalos} we show the 
  fraction of DM in minihalos in the mass range $M \in [\Mosc,\MRH]$ at $z=250$ as a function of $\TRH$ (these masses are defined in Eqs.~\ref{eq:mosc_approx} and~\ref{eq:mrh_approx}, respectively). 
  There is no collapse in $\Lambda$CDM in this mass range at this time, so all minihalos form solely 
  due to EMD. Lower $\TRH$ implies a longer period of 
  enhanced growth, while the effective pressure of lighter ALPs inhibits it. We see that 
  in the regime where the small-scale cut-off of the power spectrum is not important (the $m_a = 10^{-6}\;\eV$ curve and lower $\TRH$), 
  all of DM is already in minihalos at $z=250$. Higher $\TRH$ and lower ALP masses can result in smaller minihalo fractions. 
  Note that these calculations do not take into account possible disruption of minihalos at later times, which may liberate 
  some of the ALPs. Because the minihalos are built up hierarchically, the contents of the disrupted minihalos themselves are made of 
  smaller miniclusters (depending on the ALP mass and resulting small-scale cut-off). More detailed questions about the distribution of substructure 
 in larger ALP minihalos and galactic halos can be studied using the Extended Press-Schechter/Excursion Set formalism~\cite{Bond:1990iw,Lacey:1993iv,Zentner:2006vw}.

The formation of QCD axion miniclusters in the standard cosmological scenario with post-inflationary PQ breaking has been the topic of recent numerical simulation~\cite{Vaquero:2018tib,Buschmann:2019icd}. Ref.~\cite{Buschmann:2019icd} finds that the differential minicluster mass function peaks at $10^{-14}\msun$ with a long tail to very small masses of around $10^{-17}\msun$ and a shorter tail to heavier miniclusters up to $10^{-12}\msun$, measured at matter-radiation equality. The characteristic size of the QCD axion miniclusters in Ref.~\cite{Vaquero:2018tib} is slightly larger at $10^{-12}\msun$. Both of these studies find that the average overdensities in miniclusters are smaller than estimated in Refs.~\cite{Kolb:1993hw,Kolb:1993zz,Kolb:1994fi}. Analytic studies also based on the Press-Schechter formalism find results which are similar~\cite{Enander:2017ogx} or somewhat heavier~\cite{Fairbairn:2017sil,Fairbairn:2017dmf}. In contrast, we find that ALP miniclusters which benefit from a period of EMD are heavier than these, with masses $10^{-(6-10)}\msun$ depending on the ALP mass, as in Fig.~\ref{fig:fractional_abundance_plot}.


\subsection{Minicluster Density and Size}
\begin{figure}
  \centering
  \includegraphics[width=0.47\textwidth]{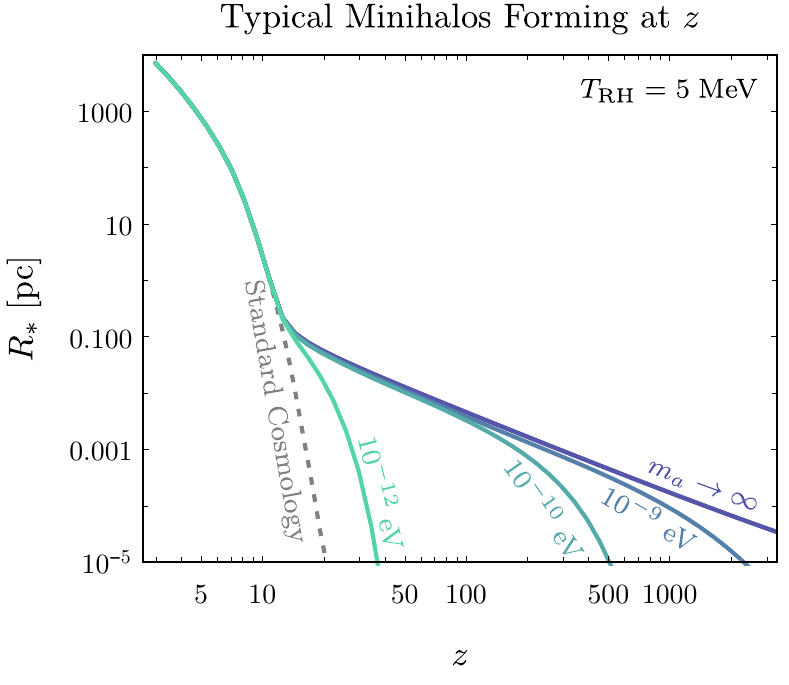}
  \caption{
    The radius of a minicluster of mass $M_*$ for $\TRH = 5\;\MeV$ and various ALP masses. 
    Color coding is the same as in the left panel of Fig.~\ref{fig:mstar_plot}.
    \label{fig:rstar_plot}
  }
\end{figure}

Collapse and decoupling from Hubble flow occur when $\delta = \delta_c \sim 1$. 
If a minicluster collapses at redshift $\zcoll$ during standard matter domination, the density 
of the final virialized object can be estimated using the spherical collapse model (see, e.g., Ref.~\cite{Galform:2010}): 
\beq
  \label{eq:collapse_density}
  \rho(\zcoll) \approx 178 \bar\rho(\zcoll) \approx 3500\;\GeV/\cm^3\; \left(\frac{1+\zcoll}{250}\right)^3.
\eeq
This allows us to calculate a characteristic radius $R_*$  of a minicluster with mass $M_*(z)$, 
\begin{align}
  R_*(\zcoll) & = \left(\frac{3M_*(\zcoll)}{4\pi\rho(\zcoll)}\right)^{1/3} \\
  & \sim 10^{-3}\;\mathrm{pc} \left(\frac{5\;\MeV}{\TRH}\right)\left(\frac{250}{1+\zcoll}\right)^{(5+n_s)/(3+n_s)},
  \end{align}
  where the second line was estimated using Eqs.~(\ref{eq:mstar_estimate}) and~(\ref{eq:collapse_density}). 
 $R_*$ is shown in Fig.~\ref{fig:rstar_plot} as function 
 of the redshift of collapse for different ALP masses. 

  Note that unlike Ref.~\cite{Nelson:2018via}, these calculations include the growth of ALP perturbations during standard matter domination after MRE. 
  As a result, we find that the miniclusters are hierarchically assembled into much more massive minihalos at later redshifts. 
  
  In the following section we discuss the implication of ALP DM clustering in the EMD 
  cosmology on direct and indirect probes.

\section{Analysis}
\label{sec:analysis}

Early matter domination enhances the growth of structure of ALP DM over a range of scales. As a result the DM distribution 
is clumpy, and the minihalo spectrum reflects fundamental properties of the ALPs and cosmology (the ALP mass and reheating temperature), 
as well as the tidal encounter history. Larger minihalos are assembled from smaller ones, but tidal disruption of the 
larger halos can ``free'' some of the component sub-halos. Because the EMD-enhanced scales collapse early, all of DM is expected to be in minihalos by the time of galaxy formation.
This observation 
has important implications for direct and indirect searches for ALP DM. First, the clumpiness of DM 
typically decreases the encounter rate of DM objects with laboratories on Earth, reducing the probability 
of a signal. On the other hand, if we are lucky and such an encounter occurs, the signal is much stronger, 
since the minihalo density is much larger than the average local DM density. 
The DM substructure also opens up a range of other probes, e.g., through pulsar timing or through gravitational lensing, as we discuss 
below.

\subsection{Minicluster Survival}
\label{sec:minicluster_survival}
Hierarchical structure formation assembled miniclusters into larger and larger objects. Both the early-time assembly 
and late-time encounters with other minihalos and dense baryonic objects (such as stars in the galactic disk) can disrupt miniclusters. 
The precise nature of the substructure in our vicinity must be modeled numerically, but we can get a sense 
of which objects survive using simple estimates, following Refs.~\cite{Goerdt:2006hp,Tinyakov:2015cgg}. 
Evolution of DM substructure has been extensively studied in the context of CDM and axion-like particles. 
Reviews of these subjects include Refs.~\cite{Koushiappas:2009du, Berezinsky:2014wya}.

\paragraph{Disruption by other miniclusters} 
Whether the EMD-enhanced minihalos at a given redshift themselves have substructure depends on the precise form 
of the power spectrum, as well as the age of the clumps. If the power-spectrum features an 
isolated peak (which is the case if $\kosc \sim \krh$), then numerical simulations of Ref.~\cite{Delos:2018ueo} indicate that 
the minihalos lack structure as one would naively expect; these objects then evolve in isolation until they are 
assembled into galaxies.  
In the other limit $\krh \ll \kosc$, a wide range of scales is enhanced by EMD, and clumps are formed from smaller clumps. Such an 
initial power spectrum was also studied in Ref.~\cite{Delos:2017thv,Delos:2018ueo}, and their results suggest that minihalos 
retain their substructure for at least a factor of $\sim 10$ in redshift after formation. 
In both cases, the analytical arguments of Ref.~\cite{vandenBosch:2017ynq} imply that tidal heating and 
stripping through minihalo-minihalo encounters is a subdominant effect to stellar and galactic disk encounters.
We therefore neglect this effect in what follows, but note that a definitive confirmation of this approximation would require an 
extremely high-resolution simulation over $\sim 10$ Gyr timescales.

Minihalo-minihalo encounters at early times are, however, important for determining the internal density 
profile $\rho(r)$ of clumps. Refs.~\cite{Delos:2017thv,Delos:2018ueo} found that $\rho(r)$ depends on whether 
the minihalos evolved in isolation (corresponding to the power-spectrum spike mentioned above) or continually 
accrete other clumps (corresponding to a scale-invariant enhancement). The latter case is analogous 
to standard hierarchical structure formation in CDM (albeit at much 
smaller scales) and results in a Navarro-Frenk-White (NFW) density profile:
\beq
\rho(r) = \frac{4\rho_s}{(r/r_s)(1 + r/r_s)^2},
\label{eq:nfw_profile}
\eeq
where $r_s$ is the scale radius and $\rho_s = \rho(r_s)$ is the scale density. 
The scale quantities, including $M_s$ (the mass interior to $r_s$), can be related 
to $R_*$ and $M_*$ used previously, and to virial quantities at any redshift. For reference, we catalog these relationships in App.~\ref{sec:mass_def_conversions}.
The spiked initial power spectrum leads to a steeper inner slope, $\rho(r) \propto r^{-3/2}$ for $r < r_s$. 
The EMD effective power spectrum in Fig.~\ref{fig:power_spectrum} is likely to result in minihalos 
with density profiles that interpolate between these two limits. In what follows, we 
take the minihalos to have the NFW profile in Eq.~\ref{eq:nfw_profile}. This assumption is conservative 
for the gravitational probes we consider in Sec.~\ref{sec:lensing_and_pta}, since a shallower 
inner slope leads to more diffuse structures, thus weakening observational prospects.

\paragraph{Disruption in stellar encounters}
If a minicluster survives structure formation, it still might be disrupted within the galaxy. 
Compared to CDM minihalos, the EMD-enhanced substructures form earlier, resulting in denser, more 
compact objects, which have a higher probability of surviving tidal stripping.
We estimate the disruption probability following Refs.~\cite{Goerdt:2006hp,Tinyakov:2015cgg}. 
An encounter with a star transfers energy to the sub-components of the minicluster, effectively 
heating them and decreasing their binding energy. Collisions with an impact parameter $b < b_c$ 
transfer enough energy to completely unbind the minicluster, where $b_c$ is the critical impact parameter~\cite{Goerdt:2006hp, Tinyakov:2015cgg},
\beq
b^2_c \sim 
\frac{G m_s \rvir}{\vrel \vvir}.
\eeq
Here $m_s$ is the typical stellar mass, 
$\vrel$ is the relative velocity of the star and minicluster, and $\rvir$ and $\vvir = \sqrt{G \Mvir/\rvir}$ are the virial radius and velocity of the minicluster -- 
these quantities can be related to the NFW scale parameters as described in App.~\ref{sec:mass_def_conversions}.
Collisions with $b>b_c$ transfer less energy but are more likely. Integrating over the entire range of impact parameters 
yields the approximate probability of disruption in a single traversal of the disk of 
$p=2\pi b_c^2 S$, where $S$ is the orbit-averaged column mass density of stars along the minicluster orbit. 
A typical minicluster has experienced $n_{\rm cross}\sim 100$ crossings in the age of the Milky Way, leading to a 
total disruption probability of~\cite{Goerdt:2006hp,Tinyakov:2015cgg} 
\begin{align}
  p & = \frac{2\pi n_{\rm cross} G \rvir S }{\vrel \vvir} \\
  & \approx \left(\frac{n_{\rm cross}}{100}\right)\left(\frac{250}{1+\zcoll}\right)^{3/2}.
  \label{eq:disruption_probability_estimate}
\end{align}
Here it is assumed that $\vrel = 10^{-3}$,  the disc has a constant stellar density, and the distribution of 
minicluster orbits is isotropic, leading to $S \approx 140 \msun/\pc^2$~\cite{Tinyakov:2015cgg}. 
Note that since this is an estimate for the disruption probability of an individual clump, it 
only depends on its density (through $1+z_c$ and Eq.~\ref{eq:collapse_density}), and it is not sensitive 
to other model details such as $m_a$ and $\TRH$ which determine the abundances of clumps of various sizes.
Ref.~\cite{Dokuchaev:2017psd} has improved on these approximations by carefully modeling the 
disc and considering interactions with halo and bulge stars. However, the total 
disruption probability remains numerically the same. Therefore we see that 
miniclusters that have formed after $z \sim 250$ are expected to have been disrupted. 
Conversely, less than 2\% of halos that formed around MRE have been 
disrupted. These numbers should be considered as guidelines rather than hard boundaries for the destruction or 
survival of substructure.
We note that high-resolution simulations of subhalos indicate that even if 
tidal encounters transfer energy far in excess of the subhalo binding energy, 
the subhalo is never completely disrupted, even for CDM subhalos~\cite{vandenBosch:2017ynq,vandenBosch:2018tyt}. 
This is because the energy injected is not efficiently redistributed among minihalo particles and therefore 
it is not directly correlated with minihalo survival.
Moreover, EMD-enhanced clumps are more compact and have an even larger chance 
of withstanding such encounters. Thus, the above estimate is most likely conservative and a more realistic  calculation of the 
DM substructure distribution today requires a simulation.

Encounters with baryonic objects can also alter the density profiles of minihalos, recently studied with $N$-body simulation in Ref.~\cite{Delos:2019tsl}. These events have a dramatic effect 
  on the minihalo density profiles at $r>r_s$, efficiently stripping away the outer minihalos, and resulting in a much 
  steeper falloff for $r>r_s$. The inner core $r < r_s$ remains NFW-like ($\rho \propto 1/r$), with 
  scale radius and density modified by $\mathcal{O}(1)$ in the encounter. The gravitational probes we consider in Sec.~\ref{sec:lensing_and_pta} 
  are mainly sensitive to the ``core'' mass $M_s$, so this modification of the density profile does not qualitatively 
  affect our results in the following sections, and we continue to use the NFW profile for simplicity.

\subsection{Direct Detection}
\begin{figure*}
  \includegraphics[width=0.47\textwidth]{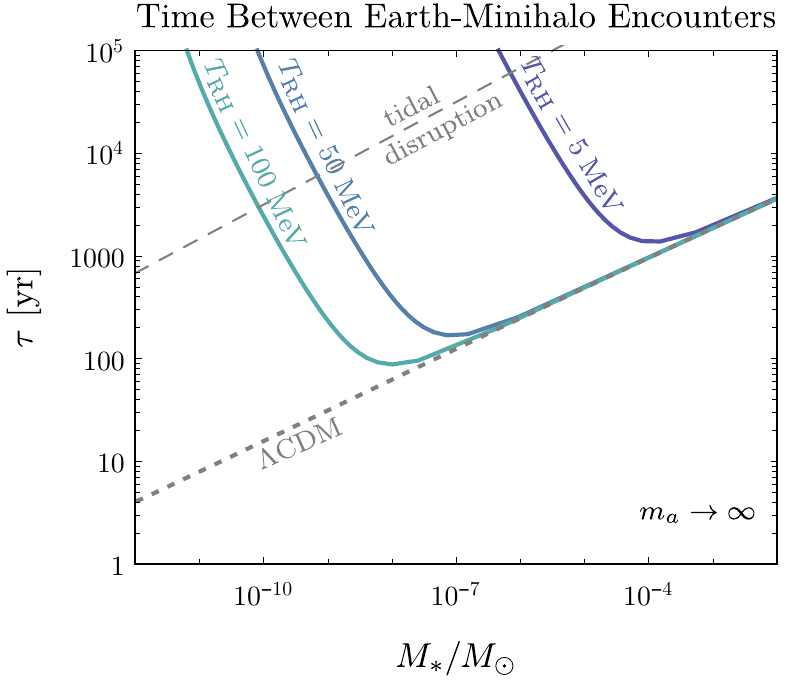}
  \includegraphics[width=0.47\textwidth]{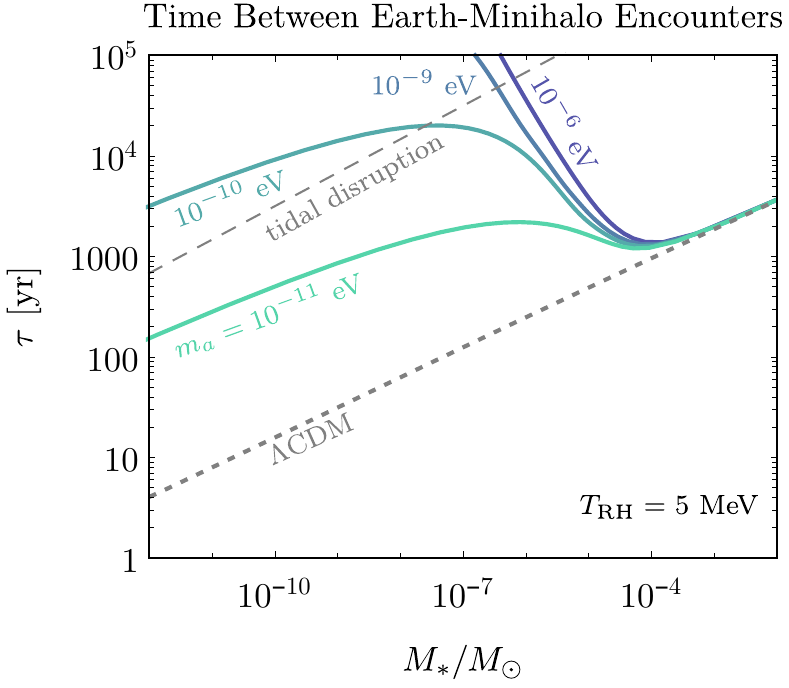}
  \caption{
    Time between Earth-minihalo encounters assuming all of DM is inside  minihalos of a single mass that survive tidal disruption until today.
    In the left panel we fix $m_a \gtrsim 10^{-6}\;\eV$ so that the small-scale cut-off due to the ALP effective pressure is irrelevant; in right 
    panel we take $\TRH = 5\;\MeV$ and vary the ALP mass. Early matter domination produces minihalos at high redshift, leading to dense and therefore 
    compact minihalos. The resulting reduced geometric cross-section increases the time between encounters for $M < \MRH$, 
    despite the increasing number density for smaller $M$. 
    Smaller ALP masses suppress growth of small scales, leading to the formation of more diffuse objects with larger cross-sections and encounter rates. 
    Gray dotted lines correspond to encounter rates for $\lcdm$ minihalos. Above the thin gray dashed line, tidal disruption of 
    minihalos due to stellar encounters is expected to be unimportant.
    \label{fig:time_between_encounters}}
\end{figure*}

Dark matter substructure has important implications for direct searches for ALPs. Direct detection experiments in terrestrial laboratories will 
detect ALPs only when Earth encounters a clump or its remnant. If the clumps remain intact, the typical time between such encounters, $\tauenc$, is 
\beq
\tauenc = \frac{1}{n \sigma \vrel},
\label{eq:time_between_encounters}
\eeq
where, assuming all of DM is in minihalos of similar mass and size, 
$n = \rho_{\mathrm{dm}}/M_*$ is the local clump number density, $\sigma \sim \pi R_*^2$ is their 
geometric cross-section and $\vrel$ velocity relative to Earth. 
Since $\sigma\propto M^{(5+n_s)/3}$, the time between encounters is smaller 
for heavier miniclusters. This is a consequence of the fact that the heavier miniclusters are less dense (due to their later formation), 
leading to a cross-section that grows faster than the constant density expectation $\propto M^{2/3}$; this effect 
compensates for the decreasing number density of heavier miniclusters at fixed local DM density $\rho_{\mathrm{dm}}$.
As we showed previously, in cosmologies with EMD, the smallest bound objects are 
assembled into larger miniclusters. Further evolution may disrupt these, so their final 
mass and size distribution depends on the merger history of the Milky Way halo and interactions with the disk. 
With this caveat in mind, we show the time between Earth-minihalo encounters in 
Fig.~\ref{fig:time_between_encounters}. The left panel 
corresponds to the CDM case with various values of $\TRH$, where ALP effective pressure effects are not important. Note that 
regime is reached already for $m_a \gtrsim 10^{-6}\;\eV$ for scales of interest. In the right panel 
we fix $\TRH = 5\;\MeV$ and show the effect of decreasing $m_a$, which suppresses the growth of structure at small scales, 
delaying collapse and leading to clumps that are more diffuse at a given mass. Their larger size leads to larger cross-sections 
and therefore higher encounter rates compared to the $m_a \rightarrow \infty$ case. 

A minihalo-Earth encounter implies a higher density of DM in the laboratory than expected from the local volume average 
and a different velocity dispersion compared to the unclustered scenario. The latter fact means that signal frequency linewidth 
$\delta f /f$ in resonant detectors is orders of magnitude smaller. The naive expectation from the typical 
kinetic energy of ALPs is $\delta f/f \approx v_s \vrel$, where $v_s^2 \sim G M_s/r_s$ is the scale velocity and $\vrel~\sim 10^{-3}$ is the relative 
Earth-minihalo velocity. 
For example, if the effective pressure effects are not important, Eqs.~\ref{eq:mstar_estimate} and~\ref{eq:collapse_density} 
imply $v_s \sim 10^{-8} (5\;\MeV/\TRH)$ (we approximated $n_s \approx 1$ and dropped factors of $g_*(\TRH)$), which is 
much smaller than the galactic virial velocity of $\sim 10^{-3}$. This suggests that the signal can be orders of magnitude narrower than 
in unclustered models. However, as the minihalo enters the solar system, it will experience 
tidal forces from the Sun, which impart different velocities to different parts of the minihalo, resulting in an 
additional drift in the signal frequency as the minihalo crosses the laboratory~\cite{Arvanitaki:2019rax}. 
The measurement time can be limited to ensure that the drift can be ignored or the drift can be incorporated as part of the signal template. 
In the former case, Ref.~\cite{Arvanitaki:2019rax} estimates that 
the minimum signal width is $\delta f/f \sim 10^{-11} (\mu\eV/m_a)^{1/2}$, which can be comparable or larger than the 
intrinsic minihalo dispersion (depending on $m_a$ and $\TRH$). As emphasized in that work, the narrow width of the signal 
in models with ALP substructure suggests that a broadband search strategy is beneficial in efficiently probing the ALP parameter space.

The previous discussion is rather optimistic, since Fig.~\ref{fig:time_between_encounters} indicates that in the absence of disruption 
the expected encounter rates in EMD cosmologies with low $\TRH$ are small. 
However, tidal disruption through interactions with the disk or other clumps can lead to a population of 
DM streams, which occupy a more significant fraction of the local volume at the price of reducing the 
density enhancement compared to an intact clump~\cite{Schneider:2010jr,Tinyakov:2015cgg,Dokuchaev:2017psd,OHare:2017yze}.
In Fig.~\ref{fig:time_between_encounters}, the region of parameter space below the thin dashed line is where tidal disruption due to interactions 
with stars may be important; this line scales as $M^{1/3}$ and corresponds to the time between encounters, Eq.~\ref{eq:time_between_encounters}, evaluated 
at fixed disruption probability as obtained from Eq.~\ref{eq:disruption_probability_estimate}. 
Thus in some parts of parameter space minihalos are expected to undergo disruption in the galactic 
gravitational field and form tidal streams. We have estimated the stream-crossing rate following Refs.~\cite{Tinyakov:2015cgg,Dokuchaev:2017psd}, 
finding that it can approach once-per-year at the price of loosing the density enhancement of the progenitor 
minihalo. However, the stream may still retain a lower velocity dispersion which can have an impact on direct detection as discussed above. 
We leave a detailed study of stream formation and properties, including the effects of a realistic minihalo distribution to future work. 

\subsection{Lensing and Pulsar Timing}
\label{sec:lensing_and_pta}

The DM clumps formed after EMD are much denser than minihalos in $\lcdm$, but are far from compact. 
The size of these miniclusters is much larger than the typical Einstein radii of stellar and supernova gravitational lens systems~\cite{Dror:2019twh}. 
As result, these objects do not lead to signals in microlensing searches such as Refs.~\cite{Tisserand:2006zx,Griest:2013aaa,Niikura:2017zjd}. 
However, new techniques to search for more diffuse objects have recently been proposed in Refs.~\cite{Dror:2019twh,Dai:2019lud,VanTilburg:2018ykj}.
We estimate the sensitivity of photometric microlensing~\cite{Dai:2019lud} and pulsar timing~\cite{Dror:2019twh} to axion minihalos 
produced from EMD below (the astrometry proposal of Ref.~\cite{VanTilburg:2018ykj} is sensitive to heavier sub-halos that cannot arise from EMD).

Photometric monitoring of individual stars behind galaxy cluster lenses can be used to observe the imprints of 
substructure on microlensing light-curves~\cite{Dai:2019lud}. The stars of interest are those which are very near to a 
cluster lens caustic; these stars are therefore highly magnified. If such a star then undergoes a microlensing event due to compact objects inside the cluster (stars or black holes), their brightness becomes variable on the time scale of the observation and the individual star can be studied~\cite{Miralda1991}. 
The lensed stars can experience magnifications of up to $\mu\sim 10^{3-4}$ under these circumstances. 
The proximity to the caustic and the resulting large amplification means that the magnification matrix 
is nearly singular -- its determinant is tuned close to zero at the level of $\mu^{-1}$. This 
tuning means that the brightness of the observed star is sensitive to surface density fluctuations in the cluster lens at 
that level. Thus, the presence of DM clumps within the cluster can then lead to $\mathcal{O}(1)$ 
brightness fluctuations if they produce surface density fluctuations of $\mathcal{O}(\mu^{-1})$ on the relevant time and length scales~\cite{Dai:2019lud}.
This phenomenon has recently been observed for a small number of stars~\cite{Kelly:2017fps,Chen:2019ncy} using the Hubble Space Telescope. Those initial measurements were not sensitive to the presence of miniclusters, which would require dedicated monitoring over a period of days or weeks either with HST or with future telescopes such as JWST.

To estimate how sensitive this technique would be to the presence of ALP minicluster we follow the simplified 
analysis of Ref.~\cite{Arvanitaki:2019rax}, based on the proposal of Ref.~\cite{Dai:2019lud}. 
We consider the miniclusters to have a standard NFW density profile (Eq.~\ref{eq:nfw_profile}), 
with a distribution described by a fractional halo mass function $df/d\ln M$. 
Assuming there is a large number of DM minihalos in the cluster along the line of sight, the resulting random 
surface density fluctuations can be described by the lensing convergence power spectrum~\cite{Dai:2019lud}, given by
\begin{equation}
  P_{\kappa}(q) = \frac{\Sigma_{\rm{cl}}}{\Sigma_{\rm{cr}}^2} \int\,dM \frac{df}{d\ln M} \left( \frac{|\tilde{\rho}(q,M)|}{M} \right)^2 \, ,
  \label{eq:lensconv}
\end{equation}
where $\Sigma_{\rm{cl}}$ is the cluster surface density, $\Sigma_{\rm{cr}}$ is the critical surface density, and $\tilde{\rho}$ is the Fourier transform of the NFW density profile, and $q = 2\pi/r$ is the inverse length scale of the fluctuations. The critical surface density can be expressed in terms of an effective distance $D_{\rm eff}$ by $\Sigma_{\rm{cr}}=1/(4\pi G D_{\rm{eff}})$. Following 
Ref.~\cite{Arvanitaki:2019rax}, we make the simplifying assumption that some fraction $f$ of the DM is contained entirely within clumps with mass $M=M_s$, so that
\begin{equation}
\frac{df}{d\ln M} = fM\delta(M-M_{s}) \,.
\end{equation}
Note that we do not use the Press-Schechter estimate of $df/d\ln M$ derived in Sec.~\ref{sec:halo_function}, 
since it does not account for minihalo disruption or sub-substructure.

\begin{figure}
  \centering
  \includegraphics[width=0.47\textwidth]{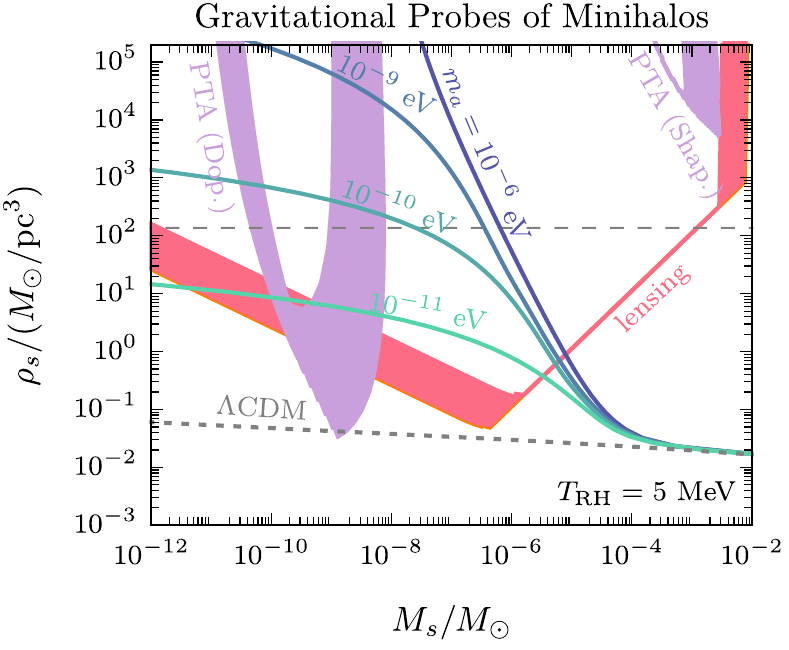}
  \includegraphics[width=0.47\textwidth]{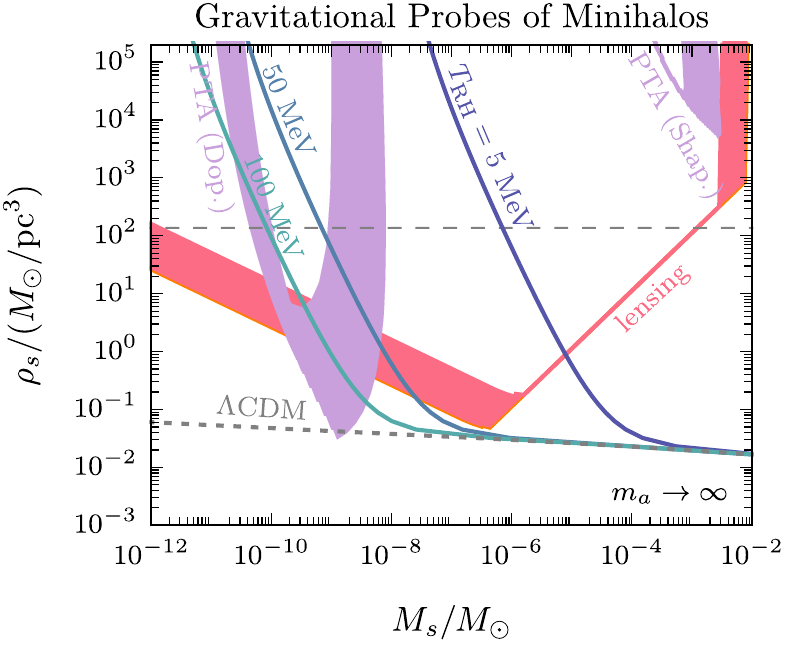}
  \caption{The reach of future pulsar timing array (PTA) Doppler and Shapiro dynamic measurements (purple) and photometric microlensing (red) in the $M_s$--$\rho_s$ plane. In each panel, the (upper) grey dashed line corresponds to a collapse redshift of 250: the region of the plane above this has $z_c>250$ with minihalos that are likely to survive tidal disruption in stellar encounters. The (lower) grey dotted line shows the prediction from the standard $\Lambda\rm{CDM}$ scenario. The left-hand panel shows EMD predictions for $m_a=10^{-6}$, $10^{-9}$, $10^{-10}$ and $10^{-11}$~eV and fixed $\TRH=5$~MeV. The right-hand panel shows EMD predictions for $\TRH=10$, $50$ and $100$~MeV for fixed $m_a\to \infty$ (this limit is already reached for $m_a \gtrsim 10^{-6}\;\eV$). The thickness of the PTA and lensing projections corresponds to varying the DM fraction in minihalos of mass $M_s$ between $0.3$ and $1$. The actual fraction of ALPs in minihalos can span a wide range and depends on model parameters (see Fig.~\ref{fig:fraction_in_minihalos}) and tidal disruption history.
\label{fig:ptaplot}}
\end{figure}

We obtain the sensitivity of photometric microlensing observations by comparing the dimensionless power spectrum $\Delta_{\kappa}(q)$ to the (expected) amplitude of observable surface density fluctuations $\mathcal{O}(10^{-3-4})$ for realistic lenses. The power spectrum is
\begin{equation}
  \Delta_\kappa (q) = \left[\frac{q^2 P_{\kappa}(q)}{2\pi} \right]^{1/2} = \frac{1}{\ln (2/\sqrt{e})}\frac{\left[ \Sigma_{\rm{cl}} f M_s \right]^{1/2} q r_s g(q r_s)}{\Sigma_{\rm{cr}}r_s}
\label{eq:dimless_convergence_ps}
  \end{equation}
where $r_s$ is the NFW scale radius, $M_s$ is the mass within the scale radius, and 
\beq
g(x)=\frac{1}{2}\sin(x)(\pi-2\Si x)-\cos(x)\Ci x
\eeq
comes from the Fourier transform of the halo density profile. The quantity $q r_s g(q r_s)$ is maximized at $qr_s=0.77$ with a value of $0.35$. We take $D_{\rm{eff}}=1$~Gpc and $\Sigma_{\rm{cl}}=0.8\Sigma_{\rm{cr}}$ -- these numbers 
roughly correspond to the observed highly-magnified star LS1~\cite{Kelly:2017fps}. 
We define the sensitivity of photometric lensing by requiring $\Delta_\kappa(q)$ in Eq.~\ref{eq:dimless_convergence_ps} to be larger than $10^{-3}$, while a number of other conditions are simultaneously satisfied. First, the length scale of the fluctuations $r=2\pi/q$ must be larger than the minimum length scale $\sim 10$~AU probed in the lens plane; this sets a lower bound in $M_s$ for the sensitivity, since smaller minihalos would give 
density fluctuations that are too rapid to be detected. Second, the characteristic size of the clump must be smaller than the largest length scale of the microlensing event, $r_s < d$ where $d\sim 10^3$~AU; minihalos that exceed this size do not give rise to star magnification fluctuations on the time scale of 
the lensing event. Note that this condition does not depend on the fraction of DM in minihalos.
Finally, to ensure that there are many clumps along the line of sight we require $f \pi (d/2)^2 \Sigma_{\rm{cl}}/M_s >10$ such that the fluctuations can 
be described by a power spectrum~\cite{Arvanitaki:2019rax}. 
This sets the sharp cut-off of $M_s\sim 10^{-2}M_{\odot}$.
To obtain the sensitivity of the future lensing search in the $M_s$--$\rho_s$ plane we evaluate Eq.~\ref{eq:dimless_convergence_ps} at the value of $q=2\pi/r$ that maximizes the sensitivity, subject to the constraints above.
We show the projected limits  as filled red regions in Fig.~\ref{fig:ptaplot}, where the thickness of the bands arises from varying the DM fraction in clumps 
of mass $M_s$ between $f=0.3$ and $1$ in the $M_s$--$\rho_s$ plane. DM substructures with scale densities above these bands will be testable with photometric lensing.
Fig.~\ref{fig:ptaplot} also shows the region which can be constrained by future pulsar timing array searches, which we now discuss.

Ref.~\cite{Dror:2019twh} has argued that even comparatively diffuse minihalos can be probed with pulsar-timing array (PTA) measurements. 
They consider two types of signal. The first is a Doppler-shift in the pulsar frequency as a DM clump passes near the star or Earth. The second is a Shapiro time-delay if a minihalo crosses our line-of-sight to the pulsar. Near-term facilities (and in particular the Square Kilometre Array) will be sensitive to these signals. Following~\cite{Dror:2019twh}, we assume that 73 currently known pulsars will continue to be monitored for the next 30 years, and that SKA will discover 200 more which it will measure for 20 years with 50~ns timing accuracy.
We show the sensitivity of a search for such anomalous frequency shifts in Fig.~\ref{fig:ptaplot} in purple. The projections shown here correspond to 30 years of observations with the current pulsar dataset, and 20 years of observations with the Square Kilometre Array (SKA) (\textit{i.e.} assuming that SKA starts in 10 years from the time of writing). The main sensitivity occurs at $M_s\sim 10^{-9}\msun$ for the Doppler dynamic signal, with the small spike feature at $M_{s}\sim 10^{-3}\msun$ corresponding to the Shapiro dynamic signal. Again, the width of the band corresponds to scanning over the clump fraction from $0.3$ to $1$. These projections were obtained by rescaling the projected sensitivity to primordial black holes by assuming the NFW density profile in Eq.~\ref{eq:nfw_profile} as described in Ref.~\cite{Dror:2019twh}.
The necessary conversions between the virial quantities used in Ref.~\cite{Dror:2019twh} and $M_s$ and $\rho_s$ are discussed in App.~\ref{sec:mass_def_conversions}.

  The left-hand panel of Fig.~\ref{fig:ptaplot} shows the target parameter space corresponding to $m_a=10^{-6}$, $10^{-9}$, $10^{-10}$ 
  and $10^{-11}$~eV for $\TRH=5$~MeV. In order to relate $R_*$ and $M_*$ obtained in Sec.~\ref{sec:halo_function} to 
  $M_s$, $r_s$ and $\rho_s$, we assumed a particular concentration parameter at formation, $c_* = R_*/r_s$ ($R_*$ is just the virial radius at 
  collapse -- see App.~\ref{sec:mass_def_conversions}). Based on the 
    compilation of simulated Earth-mass minihalos in $\Lambda$CDM from Ref.~\cite{Sanchez-Conde:2013yxa}, we take $c_* \approx 2$.\footnote{Ref.~\cite{Dai:2019lud} instead took $c_* = 4$, which is tuned to simulations of galaxy-sized DM halos.} Larger concentration parameters at formation imply more compact 
    halos (larger $\rho_s$ at fixed $M_s$) that are easier to detect with lensing or PTA. 
    Each point on the lines in Fig.~\ref{fig:ptaplot} corresponds to a specific value of the collapse epoch $z_c$ for the 
    clumps of that mass, since $z_c$ relates the scale mass to the scale radius (and hence scale density). 
    The horizontal grey-dashed line at $\rho_s\sim 10^{2}\msun/\mathrm{pc}^3$ corresponds to a collapse redshift of 250; according to 
    Eq.~\ref{eq:disruption_probability_estimate}, minihalos that form at this redshift or earlier are 
  likely to survive tidal disruption in the galaxy. We therefore expect EMD clumps which collapse before $z_c=250$ and where the ALP mass is larger than around $10^{-10}$~eV to be detectable by photometric microlensing, and possibly by Doppler dynamic PTA searches. This mass range was 
    indicated in the ALP mass-coupling parameter space of Fig.~\ref{fig:masterplot} as a green arrow. Finally, the grey dotted line shows the results for the standard $\Lambda\rm{CDM}$ scenario. Since collapse happens relatively late in that case, none of the CDM miniclusters would be dense enough to turn up in these searches (even without considering the fact that they are quite easily disrupted in hierarchical structure formation).
  The right-hand panel of Fig.~\ref{fig:ptaplot} shows the same projections as the left-hand panel but with the model curves corresponding to different reheating temperatures  $\TRH=5$, 50 and 100~MeV with the ALP mass fixed to the CDM limit of $m_a\to \infty$. We see that lensing and PTA will be sensitive to a wide range of cosmologies with reheat temperatures as high as $\TRH\sim 100\;\MeV$ (and possibly higher if our estimates of the minihalo survival probability are too conservative).

\section{Conclusion}
\label{sec:conclusion}

Early matter domination is a natural feature of many UV completions of the Standard Model, including
supersymmetric theories and various hidden sector models. If the cosmological history  included a period of EMD, both the relic abundance of dark matter and 
the growth of its density perturbations are modified relative to $\Lambda\rm{CDM}$. Non-thermally produced dark matter candidates, including axions  produced through misalignment, are particularly sensitive to the expansion history of the universe. In the axion case, EMD  yields a relic abundance that is independent of the axion mass and favors high $f_a$, including a QCD axion window of roughly $m_a\sim 10^{-(8-9)}$ eV and $f_a\sim 10^{15-16}$ GeV. 
 
In this work we studied the evolution of ALP density perturbations and the resulting DM sub-structure in cosmologies with 
early matter domination. During EMD 
density perturbations grow linearly with the scale factor, enhancing the density contrast on scales smaller than the horizon 
size at reheating and larger than the Jeans scale set by the effective ALP pressure. 
This enhancement of sub-structure results in early formation of ALP minihalos, and subsequently 
their hierarchical assembly into larger and larger objects. 
The high redshift of formation results in DM structures that have a typical density much larger than DM halos in the standard cosmology.
Given the constraints 
on late reheating, the largest objects that can benefit from EMD-enhanced growth have a mass of $\mathcal{O}(1-100)\;\MEarth$. 
 Since all of the DM ends up in minihalos, the direct detection rates are suppressed by the time 
between Earth-minihalo encounters. These times are longer than the timescale of typical experiments, making 
this search strategy impractical if all minihalos remain intact.
However, it is not clear whether all minihalos survive until today. Tidal disruption processes include clump-clump and clump-star 
encounters. If a significant fraction of clumps are disrupted, the encounter rates with the resulting streams 
may be significantly larger than for isolated minihalos. Simulations are required to reliably estimate the disruption probability over 
the lifetime of the galaxy and the resulting ALP volume-filling fraction. It will also be important to study higher-order terms in the potential, and the formation and impact of ALP solitons inside miniclusters.

If the minihalo survival probability is high, we find that proposed astrophysical detection techniques offer strong sensitivity. Pulsar timing measurements are sensitive 
to the Doppler shift induced by a minihalo passing close a pulsar. The enhanced small-scale structure can 
also have an observable imprint on the microlensing lightcurves of highly magnified stars. 
These observations are sensitive to a wide range of the relic-density target regions for 
different reheating temperatures and natural misalignment angles.
While both techniques 
require long-term observations on decadal timescales, it is important to note that the DM power spectrum at small scales is nearly unconstrained. 
Early matter domination provides an illustrative example of the interesting 
physics that can be imprinted on these scales. There are also other compelling possibilities leading to similar physics, 
including different modifications of the expansion history (e.g., a period of kination~\cite{Redmond:2018xty}), a running 
spectral index and gravitational particle production. Astrophysical searches for the small-scale structure of matter 
can thus provide a crucial window in the pre-nucleosynthesis universe and offer hints about the origin of DM.

\begin{acknowledgments}
  We thank Adrienne Erickcek, Albert Stebbins, Gabriela Barenboim, Dan Hooper, Nickolay Gnedin, Tom LeCompte, Jessie Shelton and Yang Bai for useful conversations, Jonathan Kozaczuk for collaboration in early stages of this work, and Felix for incisive criticism.
  We are especially grateful to Jeff Dror, Harikrishnan Ramani and Ken Van Tilburg for helpful discussions on pulsar timing and lensing.
This manuscript has been authored by Fermi Research Alliance, LLC under 
Contract No. DE-AC02-07CH11359 with the U.S. Department of Energy, Office of Science, Office of High Energy Physics. MJD is supported by the Australian Research Council. The work of PD is supported by NSF grant PHY-1719642. 
We thank the Galileo Galilei Institute for Theoretical Physics for the hospitality and the INFN for partial support during the completion of this work.
\end{acknowledgments}

\appendix
\section{Boltzmann Equations in Detail}
\label{app:boltzmann}
In this section we express the linear Boltzmann equations~\ref{eq:perturbed_boltzmann_system} in convenient dimensionless variables defined in 
Eq.~\ref{eq:dimless_vars} and using the scale factor $a$ as a time variable. This form of the Boltzmann system is easily implemented and solved 
numerically. The resulting perturbation equations are~\cite{Erickcek:2011us}
\begin{widetext}
\begin{subequations} 
\label{pertseta} 
\begin{align}
a^2 E \dels^\prime + \thest + 3a^2E \Phi^\prime &= a \gamt \Phi, \label{delsa}\\
a^2 E \thest^\prime + a E \thest+\tilk^2\Phi &=0, \label{thesa}\\
a^2 E \delr^\prime+\frac{4}{3}\thert+4a^2 E \Phi^\prime&= \frac{\rhost}{\rhort}a\gamt\left[\dels-\delr-\Phi\right], \label{delra}\\
a^2 E \thert^\prime+\tilk^2\Phi -\tilk^2\frac{\delr}{4}&= \frac{\rhost}{\rhort}a\gamt\left[\frac{3}{4}\thest-\ther\right], \label{thera}\\
  a^2 E \delm^\prime + \themt+3a^2E\Phi^\prime &= -3\cnad^2 a E \delta_a - 9\cnad^2 a^2 E^2 \themt/\kt^2 \label{delma} \, ,\\
  a^2 E \themt^\prime+ aE\themt+\tilk^2\Phi &= + 3\cnad^2 a E \themt + \kt^2 \cnad^2 \delta_a \label{thema} \, ,\\ 
\tilk^2\Phi +3aE^2\left[a^2\Phi^\prime+a\Phi\right] &= \frac{3}{2}a^2\left[\rhost\dels+\rhort\delr+\rhomt\delm\right], \label{phia}
\end{align}
\end{subequations}
\end{widetext}
where the prime denotes differentiation with respect to $a$, $E$ is the dimensionless Hubble parameter 
\beq
E^2 = \rhost + \rhort + \rhomt \, ,
\eeq
and 
\beq
\cnad^2 = \frac{k^2}{k^2 + 4 m^2_a a^2}.
\eeq
The energy densities are normalized as in Eq.~\ref{eq:dimless_vars}, while $\Gamma_\phi$, $m_a$ and $k$ are in units of $H_1$, the Hubble 
rate at an arbitrary initial time.

\subsection{Initial Conditions}
To derive initial conditions for the perturbation equations in Eq.~\ref{pertseta} we follow~\cite{Erickcek:2011us}, with some changes. First, we assume that all relevant modes are initially sub-horizon, such that $\kt \ll 1$. In this limit we can approximately solve the
perturbation equations while expanding in $\kt$. Super-horizon modes do not evolve, i.e. $\delta'_i=0$ and $\theta_i \sim \mathcal{O}(\kt^2)$. This implies that the right-hand side of Eqs.~\ref{delra} vanishes, leading to the following constraint at leading order in $\kt$ ,
\begin{equation}
  \delta_\phi - \delta_r - \Phi =0 \,.
 \end{equation}
Furthermore, Eq.~\ref{phia} can be solved for $\delta_\phi$ using the early-time background solutions in Eq.~\ref{eq:background_sys_early_sol}, giving
\begin{equation}
\delta_\phi = 2\Phi \,.  \label{init1}
\end{equation}
The two above equations in turn imply that
\begin{equation} \delta_r =\Phi \label{init2} \end{equation}
to leading order in $\kt$. 

For the ALP density perturbation, we assume that low-scale inflation or nontrivial inflationary dynamics prevents the generation of a large isocurvature mode. 
The adiabatic mode is zero before the ALP starts to oscillate~\cite{Marsh:2015xka}. 
After oscillations begin, the correct initial condition for the superhorizon density perturbations becomes adiabatic, 
\begin{equation}
\delm = \dels \, .  \label{init3}
\end{equation}
We show that this also follows from approximate solutions to the perturbed field equations in App.~\ref{sec:ic_from_eom}. 

Given the initial assumptions above, the right-hand side of Eq.~\ref{delma} also vanishes. This allows us to relate $\delm$ to $\themt$ via $\delm = -3aE\themt/\kt^2$, where we consider $\cnad^2 \sim\mathcal{O}(\kt^0)$. Substituting this into the right-hand side of Eq.~\ref{thema}, we observe that it vanishes, leaving
\begin{equation}
a^2 E \themt^\prime + aE\themt + \kt^2 \Phi=0 \, . 
  \end{equation}
During early matter domination we have $E(a)\simeq a^{-3/2}$. This is solved by
\begin{equation}
  \themt = -\frac{2}{3} \tilk^2 \sqrt{a}\Phi.  \label{init4}
\end{equation}
We obtain an identical solution
\begin{equation}
\thest = -\frac{2}{3} \tilk^2 \sqrt{a}\Phi  \label{init5}
  \end{equation}
from Eq.~\ref{thesa} with the same logic. Using this in Eq.~\ref{thera} along with $\delr \sim \Phi$ we finally obtain
\begin{equation}
\thert = -\frac{2}{3} \tilk^2 \sqrt{a}\Phi \,.  \label{init6}
  \end{equation}
The set of equations (\ref{init1}, \ref{init2}, \ref{init3}, \ref{init4}, \ref{init5}, \ref{init6}) form our set of initial conditions.

\section{ALP Initial Conditions from Field Equations}
\label{sec:ic_from_eom}
In this section we derive the initial conditions for the ALP density perturbation $\delta_a$ from the perturbed field equations. 
We will denote the ALP field by $\vp = \vp_0 + \vp_1$ to avoid confusion with the scale factor $a$, and work 
in the dimensionless variables defined below Eq.~\ref{eq:background_sys}.
The modes of interest enter the horizon during EMD when $\Hcon = 1/\sqrt{a}$ and the gravitational potentials are constant with 
$\Phi+\Psi = 0$.
In this regime, the background and perturbed field equations, Eqs.~\ref{eq:alp_eom} and~\ref{eq:alp_eom_perturbed}, 
simplify to 
\begin{align}
  \vp_0'' + \frac{5}{2a}\vp_0' + m_a^2 a \vp_0 & = 0\\
  \vp_1'' + \frac{5}{2a}\vp_1' + (k^2/a + m_a^2 a) \vp_1 & = +2 m_a^2 a \vp_0 \Phi 
  \label{eq:field_eom_sys}
\end{align}
where primes denote derivatives with respect to the scale factor. We take the initial 
conditions for a completely smooth field initially at rest with a misalignment value $\vp_i$:
\begin{align}
  \vp_0(0) = \vp_i, &\;\;\; \vp_0'(0) = 0 \\
  \vp_1(0) = 0, &\;\;\; \vp_1'(0) = 0.
\end{align}
The background equation is easily solved to give
\beq
\vp_0 = \frac{3\vp_i}{2m_a a^{3/2}} \sin \left(\frac{2}{3}m_a a^{3/2}\right).
\eeq
Next, we consider the super-horizon evolution of modes that enter the horizon after oscillations have begun, 
$k \ll \kosc \sim m_a^{1/3}$. This allows us to drop the $k^2$ term in Eq.~\ref{eq:field_eom_sys} in comparison 
to either the Hubble damping term $\propto \vp'$ or the mass term.
We then find the $\mathcal{O}(k^0)$ solution
\beq
\vp_1 \approx - \Phi \vp_i \cos\left(\frac{2}{3}m_a a^{3/2}\right) + \frac{3 \Phi \vp_i}{2m_a a^{3/2}} \sin \left(\frac{2}{3}m_a a^{3/2}\right).
\eeq
We can now construct the energy density and its perturbation from~\cite{Hu:1998kj} 
\begin{align}
  \rho_a & = \frac{1}{2a} (\vp_0')^2 + \frac{1}{2}m_a^2 \vp_0^2, \\
  \delta \rho_a & = \frac{1}{a}\left(\vp_0'\vp_1' + (\vp_0^\prime)^2 \Phi \right) + m_a^2 \vp_0 \vp_1.
\end{align}
Averaging over oscillations, we find that for super-horizon scales with $k\ll \kosc$ 
\begin{align}
  \rho_a &  \approx \frac{9 \vp_i^2}{16 a^3} \\
  \delta \rho_a & \approx \frac{9 \vp_i^2 \Phi}{8 a^3}
\end{align}
and therefore 
\beq
\delta_a \approx 2\Phi.
\eeq
at leading order in $k$. Thus, we see that even though the ALP field starts as completely homogeneous with $\delta_a = 0$, super-horizon evolution 
in the gravitational potential ensures that it locks onto the matter adiabatic mode after oscillations have begun.

\section{Fitting Functions for ALP Density Contrast Evolution}
\label{sec:fitting_function}
In this section we define the various semi-analytical fitting functions that we use to 
evaluate the fluctuation variance for a wide range of parameters, without having to 
solve for the evolution of all modes numerically from horizon entry to now. These fitting 
functions are nearly identical to those presented in Ref.~\cite{Erickcek:2011us} 
despite slightly different initial conditions, owing to the fact the ALP DM cannot be produced 
in the decays of the EMD field $\phi$. The reason for this similarity is that the late-time 
evolution of modes is insensitive to this initial condition. This is 
already evident in the approximate solution in Eq.~\ref{eq:emd_alp_approx_sol}, since the growing term quickly 
overtakes the initial value once the mode is inside the horizon. We confirmed this by 
comparing the various fitting functions below to the numerical solutions discussed in Sec.~\ref{sec:numsol}. 
The ALP fluid and CDM evolve differently at small scales, which we implement as a simple cut-off as we 
discuss below.

The differences between $\lcdm$ and EMD in the evolution of the DM density contrast are neatly 
encapsulated by the ratio
\begin{align}
  R(k) & = \frac{\delta_{a}}{\delta_c} \nonumber\\
& = 
\frac{A(k) \ln\left[\left(\frac{4}{e^3}\right)^{f_2/f_1}\frac{B(k) \aeq}{\ahor(k)}\right]}
{9.11 \ln\left[\left(\frac{4}{e^3}\right)^{f_2/f_1}\frac{0.594\sqrt{2}k}{\keq}\right]},
\label{eq:emd_rescaling}
\end{align}
where $\delta_c$ refers to the evolution of the CDM density contrast in $\lcdm$ and~\cite{Erickcek:2011us,Hu:1995en}
\begin{align}
f_1 & = 1 - 0.568f_b + 0.094 f_b^2 \\
f_2 & = 1 - 1.156f_b + 0.149 f_b^2 - 0.074f_b^3
\end{align}
with $f_b = \Omega_b/\Omega_m$ and~\cite{Erickcek:2011us}
\beq
\frac{\ahor(k)}{\aeq} \approx \frac{\keq}{\sqrt{2}k} \left[1 + \left(\frac{k}{\krh}\right)^{4.235}\right]^{-1/4.235}.
\eeq
This scaling relation was obtained by fitting this approximate form to numerical solutions 
of $k = \Hcon(\ahor)$; we confirmed the results of Ref.~\cite{Erickcek:2011us}.
The values of $A(k)$, $B(k)$ and in various limits are given by 
\beq
A = \begin{cases}
  0 & k > \kosc \\
  \frac{3}{5}\left(\frac{k}{\krh}\right)^2 & \kosc > k > \krh \\
  9.11 & k < \krh
\end{cases},\;\;\;
 B  = \begin{cases}
   e\left(\frac{\krh}{k}\right)^2 & k > \krh \\
   0.594 & k < \krh
 \end{cases}.
 \label{eq:a_b_coef_2}
\eeq
The functions $A$ and $B$ interpolate between the linear EMD growth and 
logarithmic RD evolution. Ref.~\cite{Erickcek:2011us} provides numerical functions that smoothly connect the two limits above. 
Note that we model the small-scale suppression of power due to the effective ALP mass by a hard cut-off at $k=\kosc$. 
This is an approximation as the actual fall-off is much smoother -- see the right panel of Fig.~\ref{fig:mode_amplitude}. 
However, this approximation allows for fast exploration of the clump parameter space without having to solve 
the full Boltzmann system for each $(m_a, \TRH)$. 

The density contrasts in Eq.~\ref{eq:emd_rescaling} are evaluated at matter-radiation equality, so the evolution at later times 
is captured by a scale-dependent growth function $TD_{\lcdm}(a,k)$ defined via Eqs.~\ref{eq:delta_parametrization} and~\ref{eq:sdgf_parametrization}. 
Since the density variance in Eq.~\ref{eq:PS_sigma_squared} involves an integral over all scales within the horizon, we need $TD_{\lcdm}(a,k)$ 
for a wide range of $k$ and $a$. We obtain $TD_{\lcdm}$ by stitching together solutions from the Boltzmann solver \class~\cite{Blas:2011rf,Lesgourgues:2011re} at small $k$ and the 
Eisenstein-Hu interpolating formula~\cite{Eisenstein:1997jh} at high $k$.
This matching is performed at $k/\keq \sim 10^{6}$ (where both the \class calculation and the Eisenstein-Hu formula are accurate) and 
scale factor $a_m$ and the result is then propagated forward or backward in time 
using solutions of the Meszaros equation~\cite{Hu:1995en}:
\beq
\delta_c'' + \frac{2+3y}{2y(1+y)}\delta_c' = \frac{3}{2y(1+y)} (1 - f_b)\delta_c,
\eeq
where primes denote derivatives with respect to $y=a/\aeq$ and $f_b = \Omega_b/\Omega_m$. 
We denote the DM density contrast with a subscript $c$ to emphasize that we are now discussing evolution in 
$\lcdm$ -- the EMD and ALP dynamics are encapsulated by $R(k)$ defined above.
This equation is valid before and after equality, and well after horizon entry; the two solutions $U_{1,2}$ are expressed in terms 
of hypergeometric functions in Ref.~\cite{Hu:1995en} and they can be matched onto the standard radiation-domination solution, given in Eq.~\ref{eq:approx_sol_rd}. 
As a result, the evolution of the density contrast can be factorized as 
\beq
\delta_c (a,k) = \mathcal{D}(a) \delta_c(\aeq, k),
\eeq
where 
\beq
\mathcal{D}(a) = U_1(a/\aeq) + \frac{A_1}{A_2}U_2(a/\aeq).
\eeq
The coefficients $A_{1,2}$ (obtained by matching in the RD regime such that $\mathcal{D} \rightarrow 1$ as $a/\aeq\rightarrow 0$) 
and the functional form of $U_{1,2}$ are given in Ref.~\cite{Hu:1995en}. 
The function $\mathcal{D}$ captures the linear growth during standard matter domination since 
$U_1(a/\aeq) \sim (a/\aeq)$ for $f_b = 0$ and $a/\aeq \gg 1$. The full expression is 
accurate near the transition from logarithmic growth (captured by $U_2$) to linear evolution, 
and $f_b \neq 0$.\footnote{On scales larger than the baryonic Jeans length, the $f_b = 0$ solution 
is appropriate since the baryons are no longer pressure-supported and collapse like CDM. We interpolate between the $f_b = 0$ and $f_b \neq 0$ regimes using the prescription in 
Ref.~\cite{Erickcek:2011us}.} The scale-dependent growth function at an arbitrary redshift is then 
given by
\beq
TD_{\lcdm}(a,k) = TD_{\lcdm}(a_m,k) \left[\frac{\mathcal{D}(a)}{\mathcal{D}(a_m)}\right].
\eeq
The scale factor $a_m$ at which the numerical and semi-analytic expressions for $TD_{\lcdm}$ are matched is arbitrary, 
and can be chosen to minimize the error made in the simple extrapolation using $\mathcal{D}$. Similar to Ref.~\cite{Erickcek:2011us}, 
we find that matching at $z_m=50$ and using $\mathcal{D}(a)$ leads to fractional errors of $< 4\%$ for a wide range of 
redshifts and scales of interest. This procedure ensures that the amplitude of density fluctuations on large 
scales is correctly normalized. In particular, using Eq.~\ref{eq:PS_sigma_squared}, we reproduce 
the observed value of $\sigma_8 = \sigma(z=0,R=8/h\;\Mpc) \approx 0.8$~\cite{Aghanim:2018eyx}.

\section{Isocurvature Constraints}
\label{sec:isocurvature}

Planck~\cite{Akrami:2018odb} constrains ratio of the scalar-to-isocurvature amplitude to be $\alpha   < 0.038$.
For theories with a period of EMD we proceed following Ref.~\cite{Fox:2004kb}. The definition of the isocurvature perturbation is
\begin{equation}
  S_i = \frac{\delta(n_i/s)}{n_i/s} = \frac{\delta n_i}{n_i} -3\frac{\delta T}{T} \,.
  \label{eq:isoc}
  \end{equation}
  We assume that only the axion has $S_a \neq 0$, with all other fields satisfying $S_i=0$.
  The invariance of the local energy density under isocurvature perturbations can then be used to relate $S_a$ to the temperature perturbation $\delta T/T$ as follows. 
  The total energy is
\begin{equation}
  \rho = \sum_i m_i n_i + m_a n_a + \rho_r
  \end{equation}
and we require
\begin{equation}
\delta\rho = \sum_i m_i  \delta n_i + m_a \delta n_a + 4\rho_r \frac{\delta T}{T} \simeq 0 \,.
  \end{equation}
The second equality in Eq.~\ref{eq:isoc} implies $\delta n_i = n_i 3 \delta T/T$ for $i\neq a$. Substituting that in the equation above we obtain
\begin{equation}
  \frac{\delta T}{T} \simeq - \frac{\rho_a}{3\sum_i\rho_i+4\rho_r} S_a,
  \label{eq:deltaT}
  \end{equation}
which implicitly requires $\delta T/T\simeq 0$ for the ALP. For fluctuations on super-horizon scales which enter the horizon during standard matter domination (after $T_{\rm{eq}}$) the above implies
\begin{equation}
\frac{\delta T}{T} \simeq -\frac{1}{3} \frac{\Omega_a}{\Omega_m} S_a \,. 
  \end{equation}
We add an extra  $-1/15$ onto the prefactor to take the Sachs-Wolfe effect into account~\cite{Fox:2004kb}, so the total temperature fluctuation is 
\begin{equation}
  \frac{\delta T}{T}_{\rm{iso}} \simeq -\frac{6}{15} \frac{\Omega_a}{\Omega_m} S_a \,.
\end{equation}

Now we find an expression for $S_a$ in terms of $\theta_i$, $f$ and $H_I$. 
The fractional fluctuation in the axion density is related to fluctuations around the initial misalignment angle $\theta_i$ in the early universe
\begin{equation}
S_a \simeq \frac{\delta n_a}{n_a}\simeq \frac{\delta(\theta^2)}{\langle\theta^2\rangle} \simeq \frac{\langle( \theta \rangle +\delta\theta )^2-\langle \theta^2 \rangle}{\langle \theta^2 \rangle} \simeq \frac{2\theta_i \delta\theta + (\delta\theta)^2  -\sigma_\theta^2}{\theta_i^2+\sigma_\theta^2}
  \end{equation}
  where the variance is $\sigma_\theta^2=\langle \left( \theta - \langle\theta\rangle \right)^2 \rangle = H_I^2/(2\pi f)^2$ and $\langle\theta\rangle =\theta_i$. We will also need
\begin{equation}
\langle S_a^2 \rangle  = 2\sigma_\theta^2 \frac{2\theta_i^2 + \sigma_\theta^2}{\left(\theta_i^2 +\sigma_\theta^2\right)^2} \,.
  \end{equation}

The isocurvature component of the total power in CMB temperature fluctuations is
\begin{equation}
\alpha  =\frac{\langle (\delta T/T)_{\rm iso}^2 \rangle }{\langle (\delta T/T)_{\rm tot}^2 \rangle} \, ,
  \end{equation}
where COBE measured $\langle (\delta T/T)_{\rm tot}^2 \rangle^{1/2}=1.1\times 10^{-5}$~\cite{Bennett:1996ce}. Putting the pieces together we have
\begin{equation}
  \alpha =  \left( \frac{6}{15}\right)^2 \frac{(\Omega_a/\Omega_m)^2}{\langle (\delta T/T)_{\rm tot}^2 \rangle} 2\sigma_\theta^2 \frac{2\theta_i^2 + \sigma_\theta^2}{\left(\theta_i^2 +\sigma_\theta^2\right)^2} < 0.038\,.
  \label{eq:IsoCon}
\end{equation}
where $\Omega_m \simeq 0.13$,  the expression for the axion relic density in EMD theories is Eq.~\ref{eq:Omega_EMD}  and the rest is defined above.

As an example, for benchmark values of $\theta_i=1$, $f_a=9\times10^{14}$~GeV and $\TRH=10$~MeV Eq.~\ref{eq:IsoCon} implies $H_I<  2\times10^{10}$~GeV. The requirements on the scale of inflation in EMD theories are less onerous than standard cosmology by a factor of $\mathcal{O}(10^{2-3})$. As we raise the reheating temperature the bound on the $H_I$ decreases: for $T_{\rm{RH}}=500$~MeV we have $H_I < 2\times 10^9$~GeV.
The isocurvature bounds can be evaded in low-scale theories of inflation -- see, e.g., Refs.~\cite{German:2001tz,Martin:2013tda}.

\section{Relationships Between Size and Mass Definitions}
\label{sec:mass_def_conversions}
In this section we relate the various mass and size scales used 
to characterize minihalos.
We will assume that minihalos have an NFW profile:
\beq
\rho(r) = \frac{4\rho_s}{(r/r_s)(1+r/r_s)^2},
\eeq
where $\rho_s$ and $r_s$ are scale density and scale radius. The mass within 
a certain radius $r$ is then
\beq
M(r) = 16\pi \rho_s r_s^3 f(r/r_s),
\eeq
where 
\beq
f(c) = \ln(c+1) - \frac{c}{c+1}.
\eeq
The scale mass $M_s$ is the mass within the scale radius:
\beq
M_s = M(r_s) = 16\pi \rho_s r_s^3 f(1).
\label{eq:scale_mass_def}
\eeq
Given any two of $(r_s, \rho_s, M_s)$ we can solve 
for the other one using this relationship.

\subsection{Virial mass, radius and concentration}
The virial quantities are defined for a sphere centered on the 
gravitational potential minimum that encloses a region within which 
the average density is $\Delta = 200$ times the critical density at some redshift $\rho_c(z) = 3H^2(z)/(8\pi G)$~\cite{Ludlow:2013vxa,Sanchez-Conde:2013yxa} 
(sometimes only the average matter density $\Omega_m \rho_c(z)$ is used~\cite{Goerdt:2006hp}). 
We therefore have the following relationship between $\Mvir$ and $\rvir$:
\beq
\Mvir = \frac{4\pi}{3} 200 \rho_c \rvir^3.
\eeq

We want to relate virial quantities to the NFW scale parameters defined above. First, 
\beq
\Mvir = M(\rvir) = M_s f(c_{200})/f(1),
\eeq
where we defined the concentration parameter
\beq
c_{200} = \rvir/r_s.
\eeq
We can take $(c_{200},\Mvir)$ as defining the halo and solve for the NFW scale quantities using the 
relations above, leading to 
\beq
M_s = \frac{f(1)}{f(c_{200})} \Mvir,
\label{eq:scale_mass_from_virial_mass}
\eeq
\beq
r_s = \rvir/c_{200}
\eeq
and 
\beq
\rho_s = \frac{1}{12} 200 \rho_c \frac{c_{200}^3}{f(c_{200})}
\eeq
Note that at large concentration parameters $c_{200}\gg 1$, $f(c_{200}) \sim \ln c_{200}/e$ 
and Eq.~\ref{eq:scale_mass_from_virial_mass} implies that 
\beq
\frac{\Mvir}{M_s} \sim 43 + 5 \ln \left(\frac{c_{200}}{10^4}\right),
\eeq
so the virial mass and scale mass can be quite different. This occurs if the 
redshift at which the virial quantities are calculated is long after the initial collapse that created the gravitationally bound core; 
subsequent evolution resulted in accretion of matter on this core.

\subsection{\texorpdfstring{$M_*$}{Mstar}, \texorpdfstring{$R_*$}{Rstar} and Concentration at Formation}
Recall that $R_*$ is defined as the radius within which 
the density is $178\rho_a(z_c)$, where $\rho_a(z_c)$ is the 
background density at collapse. This means that $M_*$ and $R_*$ 
are like the virial quantities, evaluated at collapse, i.e. $\Mvir(z_c) \approx M_*$ and $\rvir(z_c) \approx R_*$.
Let $c_*$ be the concentration parameter at formation, 
i.e. 
\beq
c_* = R_*/r_s.
\eeq
This is an $\mathcal{O}(1)$ number. Simulations of Earth-mass $\Lambda$CDM halos imply that $c_* \approx 2$~\cite{Sanchez-Conde:2013yxa} for 
$M_s \sim 10^{-6}\msun$, but in principle this is a cosmology and mass-dependent quantity.
The concentration parameter allows us to find $M_s$ and $\rho_s$ from $c_*$, $R_*$ and $M_*$:
\beq
M_s = \frac{f(1)}{f(c_*)} M_*
\eeq
The scale density is then obtained from Eq.~\ref{eq:scale_mass_def}.

\bibliography{biblio}
\end{document}